\documentclass[3p]{elsarticle}

\usepackage{hyperref}
\usepackage{amssymb, amsmath}
\usepackage{wasysym}
\usepackage{siunitx}
\usepackage[stable]{footmisc}

\usepackage{xcolor}

\biboptions{authoryear}

\begin{document}

\begin{frontmatter}

\title{Targeting ultra-high energy neutrinos with the ARIANNA experiment}


\author[UCI]{A. Anker}
\author[UCI]{S. W. Barwick}
\author[Uppsala2]{H. Bernhoff}
\author[Kansas,Moskov]{D. Z. Besson}
\author[Uppsala]{Nils Bingefors}
\author[UCI]{G. Gaswint}
\author[UCI]{C. Glaser\corref{mycorrespondingauthor}}
\cortext[mycorrespondingauthor]{Corresponding author}
\ead{christian.glaser@uci.edu}
\author[Uppsala]{A. Hallgren}
\author[Whittier]{J. C. Hanson}
\author[UCI,ECAP]{R. Lahmann}
\author[Kansas]{U. Latif}
\author[Taiwan]{J. Nam}
\author[Kansas,Moskov]{A. Novikov}
\author[Berkeley]{S. R. Klein}
\author[UCI2]{S. A. Kleinfelder}
\author[DESY,HU]{A. Nelles}
\author[UCI]{M. P. Paul}
\author[UCI]{C. Persichilli}
\author[UCI]{S. R. Shively}
\author[UCI,UCIC]{J. Tatar}
\author[Uppsala]{E. Unger}
\author[Taiwan]{S.-H. Wang}
\author[UCI]{G. Yodh}

\address[UCI]{Department of Physics and Astronomy, University of California, Irvine, CA 92697, USA}
\address[Uppsala2]{Uppsala University Department of Engineering Sciences, Division of Electricity, Uppsala, SE-752 37 Sweden}
\address[Kansas]{Department of Physics and Astronomy, University of Kansas, Lawrence, KS 66045, USA}
\address[Moskov]{National Research Nuclear University MEPhI (Moscow Engineering Physics Institute), Moscow 115409, Russia}
\address[Uppsala]{Uppsala University Department of Physics and Astronomy, Uppsala, SE-752
37, Sweden}
\address[Whittier]{Whittier College Department of Physics, Whittier, CA 90602, USA}
\address[ECAP]{ECAP, Friedrich-Alexander Universit\"at Erlangen-N\"urnberg, 91058 Erlangen, Germany}
\address[Taiwan]{Department of Physics and Leung Center for Cosmology and Particle Astrophysics, National Taiwan University, Taipei 10617, Taiwan}
\address[Berkeley]{Lawrence Berkeley National Laboratory, Berkeley, CA 94720, USA}
\address[UCI2]{Department of Electrical Engineering and Computer Science, University of California, Irvine, CA 92697, USA}
\address[DESY]{DESY, 15738 Zeuthen, Germany}
\address[HU]{Humbolt-Universit\"at zu Berlin, Institut f\"ur Physik, 12489 Berlin, Germany}
\address[UCIC]{Research Cyberinfrastructure Center, University of California, Irvine, CA 92697 USA}

\begin{abstract}
The measurement of ultra-high energy (UHE) neutrinos (E $>$ \SI{e16}{eV}) opens a new field of astronomy with the potential to reveal the sources of ultra-high energy cosmic rays especially if combined with observations in the electromagnetic spectrum and gravitational waves. 
The ARIANNA pilot detector explores the detection of UHE neutrinos with a surface array of independent radio detector stations in Antarctica which allows for a cost-effective instrumentation of large volumes.
Twelve stations are currently operating successfully at the Moore's Bay site (Ross Ice Shelf) in Antarctica and at the South Pole. We will review the current state of ARIANNA and its main results. We report on a newly developed wind generator that successfully operates in the harsh Antarctic conditions and powers the station for a substantial time during the dark winter months. The robust ARIANNA surface architecture, combined with environmentally friendly solar and wind power generators, can be installed at any deep ice location on the planet and operated autonomously. We report on the detector capabilities to determine the neutrino direction by reconstructing the signal arrival direction of a \SI{800}{m} deep calibration pulser, and the reconstruction of the signal polarization using the more abundant cosmic-ray air showers. 
Finally, we describe a large-scale design -- ARIA -- that capitalizes on the successful experience of the ARIANNA operation and is designed sensitive enough to discover the first UHE neutrino. 

\end{abstract}


\end{frontmatter}

\section{Introduction}
The origin of ultra-high energy cosmic rays (UHECRs) is one of the biggest mysteries in astroparticle physics. The deflection of cosmic rays in galactic and extra-galactic magnetic fields makes the determination of their origin extremely challenging, and despite decades of research, no UHECR source could be identified. A more promising way to track down the sources of cosmic rays is the measurement of ultra-high energy (UHE) neutrinos as neutrinos traverse the universe unimpeded and point back to their source. 
Neutrinos are either produced directly at an astrophysical object through the acceleration of cosmic rays and succeeding interactions with matter surrounding the source (called astrophysical neutrinos), or during the propagation of cosmic rays through the universe via interactions with photons from the cosmic microwave background (called cosmogenic neutrinos). 

Recently, the IceCube neutrino detector at the South Pole reported a detection of a \SI{3e14}{eV} astrophysical neutrino in coincidence with a flaring state of a blazar that was observed with gamma-ray telescopes \citep{IceCubeBlazer2018}. With a significance of \SI{3}{\sigma}, it constitutes the first evidence of a cosmic-ray source. This discovery shows the strength of neutrino astronomy and the multi-messenger approach as one of the best opportunities to discover the sources of cosmic rays also at even higher energies. 

As the interaction cross section of UHE neutrinos is small, huge volumes need to be instrumented to observe neutrinos with sufficient statistics. The ice sheets of Antarctica and Greenland provide a suitable detection medium, and a sparse instrumentation of the ice with sensors allows for the construction of terraton detectors.
Neutrinos with energies above \SI{e16}{eV} are measured best with the radio technique \citep{IceCubePRL2016} because of the small attenuation of radio signals in ice ($L_a \approx \SI{1}{km}$). A sparse array of radio detector stations allows for a cost-effective instrumentation of large volumes whereas optical detectors, such as IceCube, become cost-prohibitive because of the larger attenuation and scattering of optical light in ice.

\begin{figure}
    \centering
    \includegraphics[width=0.49\textwidth]{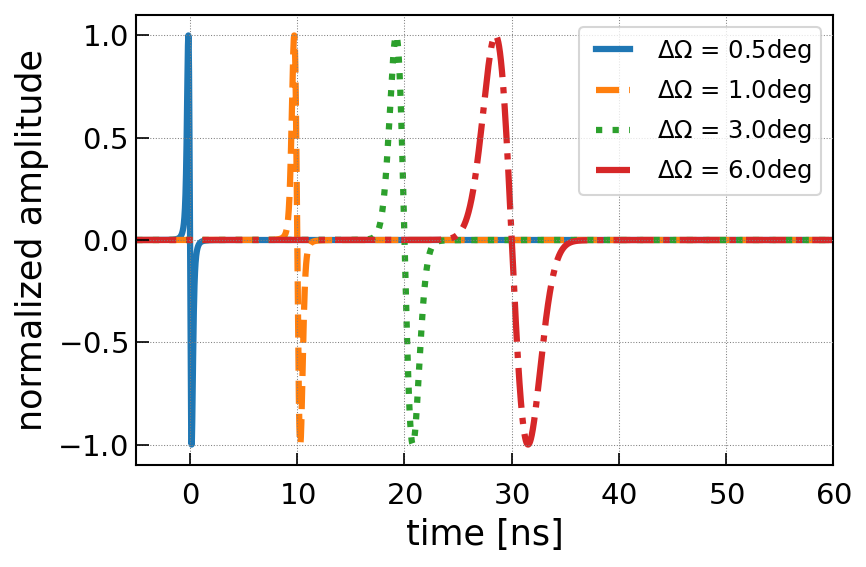}
    \includegraphics[width=0.49\textwidth]{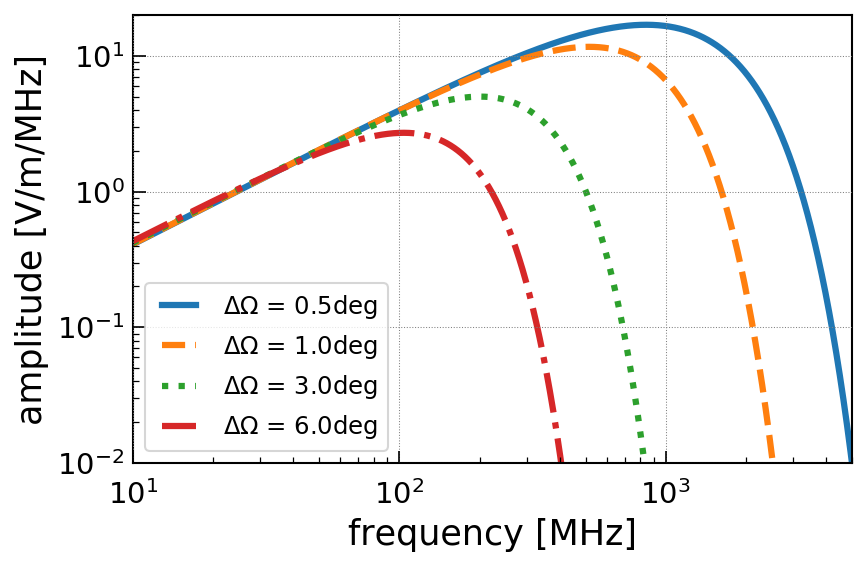}
    \caption{Askaryan radio pulse for different viewing angles with respect to the Cherenkov cone = \SI{0}{\degree} for a hadronic shower according to the model presented in \cite{AlvarezMuniz2000} and \cite{Alvarez-Muniz1998}. (left) Time domain representation. The pulse start time is shifted for better visibility. (right) Frequency spectrum.}
    \label{fig:askaryan}
\end{figure}

Radio signals in ice are generated via the Askaryan effect \citep{Askaryan1962}: a high-energy neutrino interaction induces a particle shower in the ice and a time-varying negative charge excess in the shower front leads to a short radio flash of a few nanoseconds length with relevant frequencies from \SI{50}{MHz} to \SI{1}{GHz}. The radio emission is emitted on a narrow cone -- the Cherenkov cone -- around the shower axis where the emission from all parts of the shower adds up coherently. The cone has an opening angle of approx. \SI{56}{^\circ} in ice and a width of a few degrees, as depicted in Fig.~\ref{fig:sketch}. Hence, the observable Askaryan radio pulse depends strongly on the geometry as shown in Fig.~\ref{fig:askaryan}. If the shower is observed directly on the Cherenkov cone, the signal is strongest and extends to frequencies of \SI{\sim1}{GHz}. If the shower is observed further off the cone, the cutoff frequency moves down to lower frequencies. Thus, the relevant frequency range is \SI{\sim50}{MHz} to \SI{\sim1}{GHz}, and a broadband detection is beneficial to be sensitive to all viewing angles and to measure the shape of the frequency spectrum which is crucial to extract the direction, energy and flavor of the neutrino. Another consequence of the emission on the Cherenkov cone is that both the incoming signal direction as well as the signal polarization need to be measured to determine the neutrino direction.

This promising detection technique is currently explored by two pilot arrays with slightly different designs. The ARA detector \citep{ARA} at the South Pole places antennas at a depth of \SI{200}{m} into narrow holes ($\diameter \approx \SI{15}{cm}$) that are drilled from the ice surface. This comes with the advantages of a larger sensitivity per station, and being less impacted by firn effects: In the upper part of the ice layer (at the South Pole the upper \SI{\sim200}{m}), the index of refraction changes from 1.78 of clear deep ice to 1.35 at the surface \citep{Kravchenko2004, Barwick2018} because of a change in the ice density. Therefore, the radio signals do not propagate on straight lines but undergo continuous Fresnel refraction. 
Still, the drilling is a substantial deployment effort and only antennas that fit into the narrow hole can be used which limits the broadband response and the sensitivity to the horizontal signal polarization. 

The ARIANNA detector, in turn, consists of autonomous detector stations located just slightly below the ice surface. This allows a quick deployment and the installation of large high-gain broadband LPDA antennas with different orientations, which enables a precise measurement of the signal polarization to reconstruct the neutrino direction, and the measurement of the frequency spectrum which is required to determine the neutrino energy. 
Being close to the surface comes with additional advantages: In dipole antennas a couple of meters below the surface (which is a straight-forward extension of the current station layout and part oft the ARIA concept), the Askaryan radio pulse can be observed two times, one pulse that propagates directly to the antenna and one that is reflected off the ice surface,  for almost all events. This allows for an unambiguous neutrino identification, and the reconstruction of the distance to the interaction vertex. Furthermore, an ARIANNA station is also sensitive to the radio signals of cosmic-ray air showers which allows for an in-situ calibration, and continuous test and monitoring of the detector under realistic conditions. 
However, the smaller exposure to neutrinos per station requires the installation of more stations to reach the same sensitivity as an array of deep detectors, and firn effects need to be taken into account in the data analysis.

In this paper we report on the accomplishments of the ARIANNA pilot array and present a improved design of a large-scale radio neutrino detector -- called ARIA -- that has emerged from the ARIANNA experience. We will review relevant previously published ARIANNA results for completeness and present several new achievements relevant for the design of ARIA, and other Askaryan based high-energy neutrino detectors with a significant surface component of LPDA antennas. We present a novel wind generator specifically designed to work at the harsh Antarctic conditions capable of powering the station through the dark winter months. The robust ARIANNA surface architecture, combined with environmentally friendly solar and wind power generators, can be installed at any deep ice location on the planet and operated autonomously. This design avoids the necessity to lay power and communication cables from a centralized power generator, which is expensive and geographically restrictive.

We report on an in-situ calibration measurement at the South Pole where the signal arrival direction is reconstructed from pulses of a \SI{800}{m} deep calibration pulser and discuss its implications on the understanding and modelling of signal propagation through the ice. We then discuss how neutrino radio signals can be identified and demonstrate this technique using the more abundant cosmic-ray air showers. Furthermore, air-shower radio signals are used to show the capabilities to reconstruct the signal polarization, an important quantity to determine the neutrino direction. 
Finally, we will present the array and detector layout of ARIA, briefly discuss how neutrino signals are identified and how the neutrino properties can be reconstructed, and present its expected sensitivity. 

\section{The ARIANNA detector}

\begin{figure}
    \centering
    \includegraphics[width=0.7\textwidth]{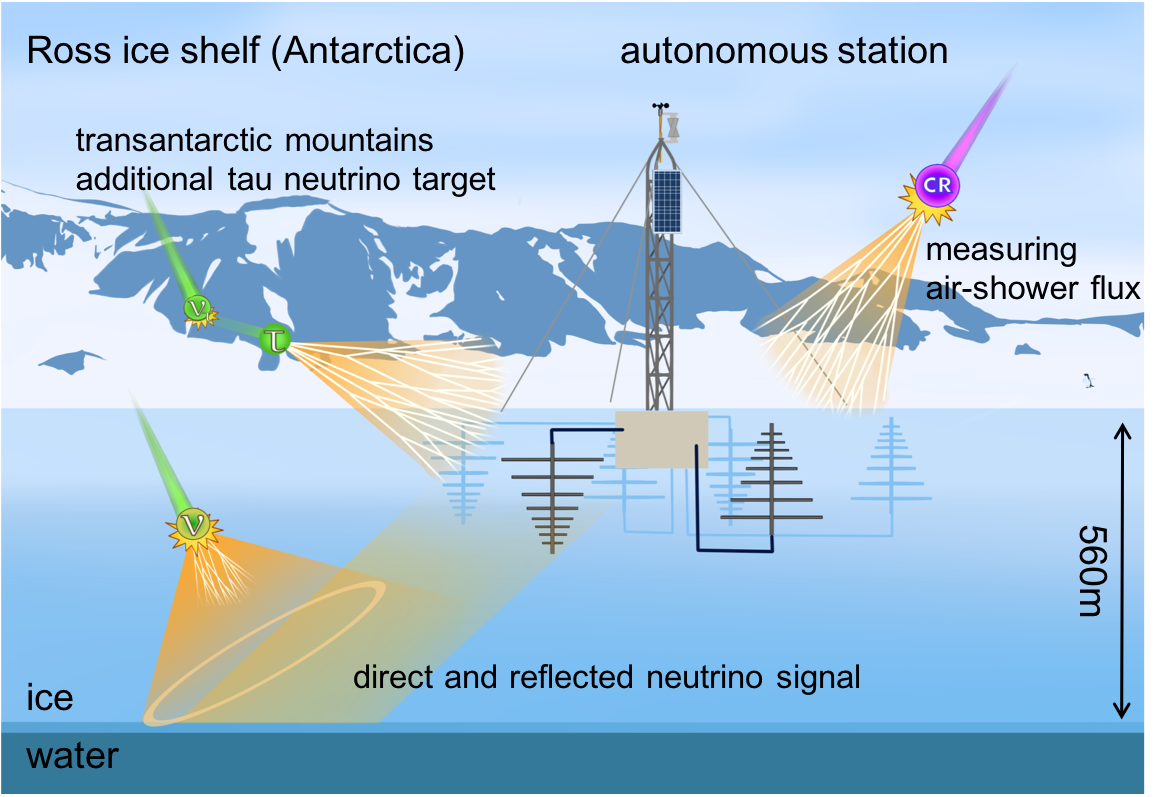}
    \caption{Sketch of the ARIANNA detector at the Ross Ice Shelf. }
    \label{fig:sketch}
\end{figure}

The ARIANNA detector consists of autonomous and independent stations, i.e., the information of one station is sufficient to measure a neutrino and multi-station coincidences are not required. The station layout is depicted in Fig.~\ref{fig:sketch}. Each station comprises two pairs of downward facing LPDA antennas with orthogonal orientation and are spatially separated by \SI{6}{m}. In the second generation of ARIANNA stations, the downward facing LPDAs for neutrino detection are complemented by two pairs of upward pointing LPDAs for cosmic-ray detection and vetoing. 

ARIANNA stations are solar powered and communication takes place via the Iridium satellite network or a high-speed long-range wifi connection. The ARIANNA pilot array was initially deployed at Moore's Bay on the Ross ice shelf near the coast of Antarctica \citep{ARIANNAPrototype2010} but the autonomous nature of the ARIANNA design allows a deployment at any suitable site around the world, so that two ARIANNA stations were deployed in 2017 and 2018 at the South Pole and have been operating successfully since then. 

The main part of the ARIANNA pilot array is the hexagonal radio array (HRA) \citep{Barwick2014}. It consists of seven 4-channel stations and has been installed at the Moore's Bay site. All stations have been operating successfully since their deployment (the first stations were deployed in 2012) demonstrating the enormous stability of the ARIANNA hardware in harsh Antarctic conditions. 

The ARIANNA hardware is based on the SST chip design \citep{Kleinfelder2015}. It currently supports up to 8 input channels on a single board. The input is sampled at \SI{1}{GSPS} or \SI{2}{GSPS} and continuously stored in a switched capacitor array. The SST chip has a precise time synchronization between samples and across channels of less than \SI{5}{ps} \citep{PhDEdwin} which allows for a precise reconstruction of the signal direction despite the small spatial extent of an ARIANNA station. Following a trigger, 256 samples are digitized with 12 bit ADCs and read into an FPGA and thereafter into an Mbed micro processor \citep{Mbed} for the calculation of a second trigger stage and data storage. 

Interesting events are triggered using a high and low threshold crossing requirement: The input signal needs to cross a remotely adjustable positive and negative threshold within \SI{5}{ns}. This requirement substantially reduces the trigger rate on thermal noise fluctuations compared to a single threshold trigger while retaining the same trigger efficiency on Askaryan pulses. To further reduce the trigger rate on thermal noise fluctuations, coincidences between multiple channels are required. A typical setting is to require 2 triggers out of the 4 channels within \SI{30}{ns} which corresponds to the maximal physical allowed propagation time between the antennas. The thresholds are set to 4 times the RMS noise and results in a typical trigger rate of \SI{10}{mHz}. The threshold are typically set only once at the beginning of the season because the noise conditions are very stable and a continuous adjustment of thresholds is not required. We note that we observe short periods of elevated rates caused by non-thermal noise, i.e., the elevated rates are not caused by a change in the average RMS noise. Most of these occurrences are caused by high-wind speeds.

We occasionally observe continuous narrowband signals from, e.g., air plane communication or weather balloons. We therefore compute an additional L1 trigger that rejects events with a strong signal in only one frequency bin. More advanced trigger schemes are not necessary at this stage of the experiment given the extremely radio quiet environment at Moore's Bay and at the South Pole. 

\subsection{Overview of ARIANNA stations}
\begin{figure}
    \centering
    \includegraphics[width=0.4\textwidth]{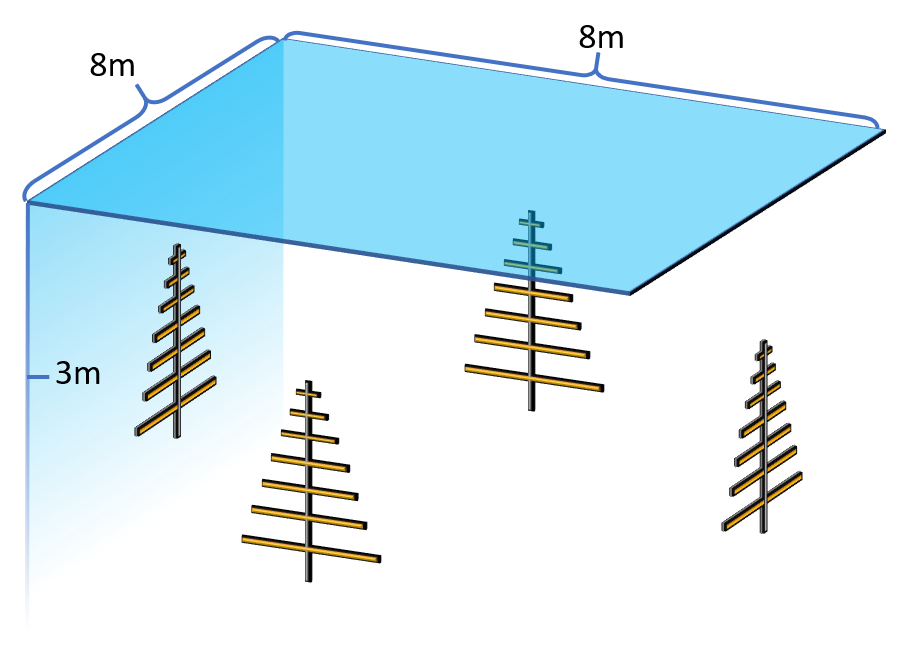}
    \includegraphics[width=0.4\textwidth]{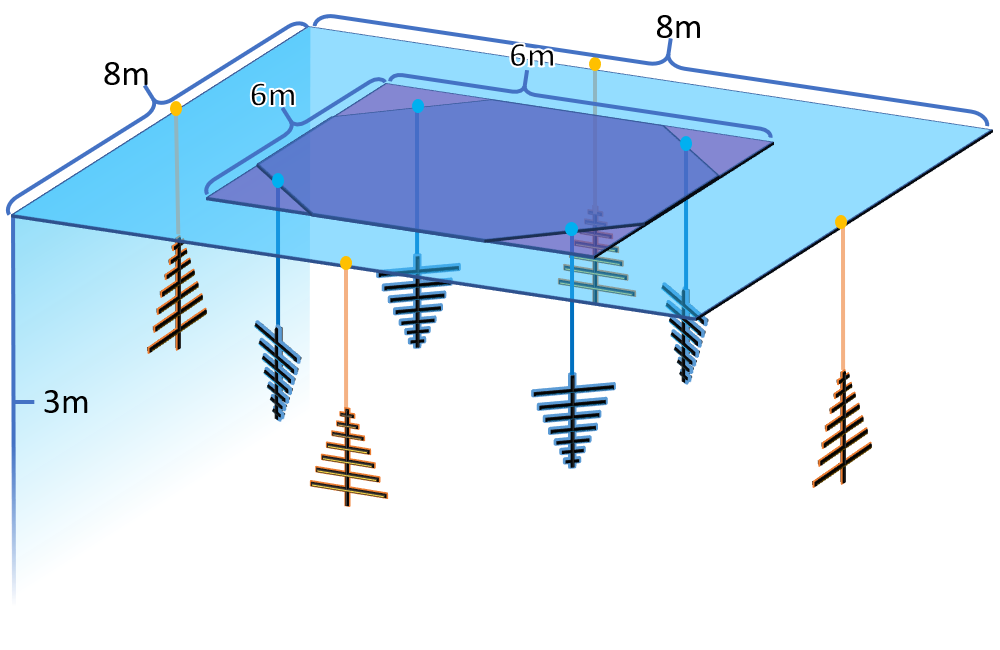}
    \includegraphics[width=0.4\textwidth]{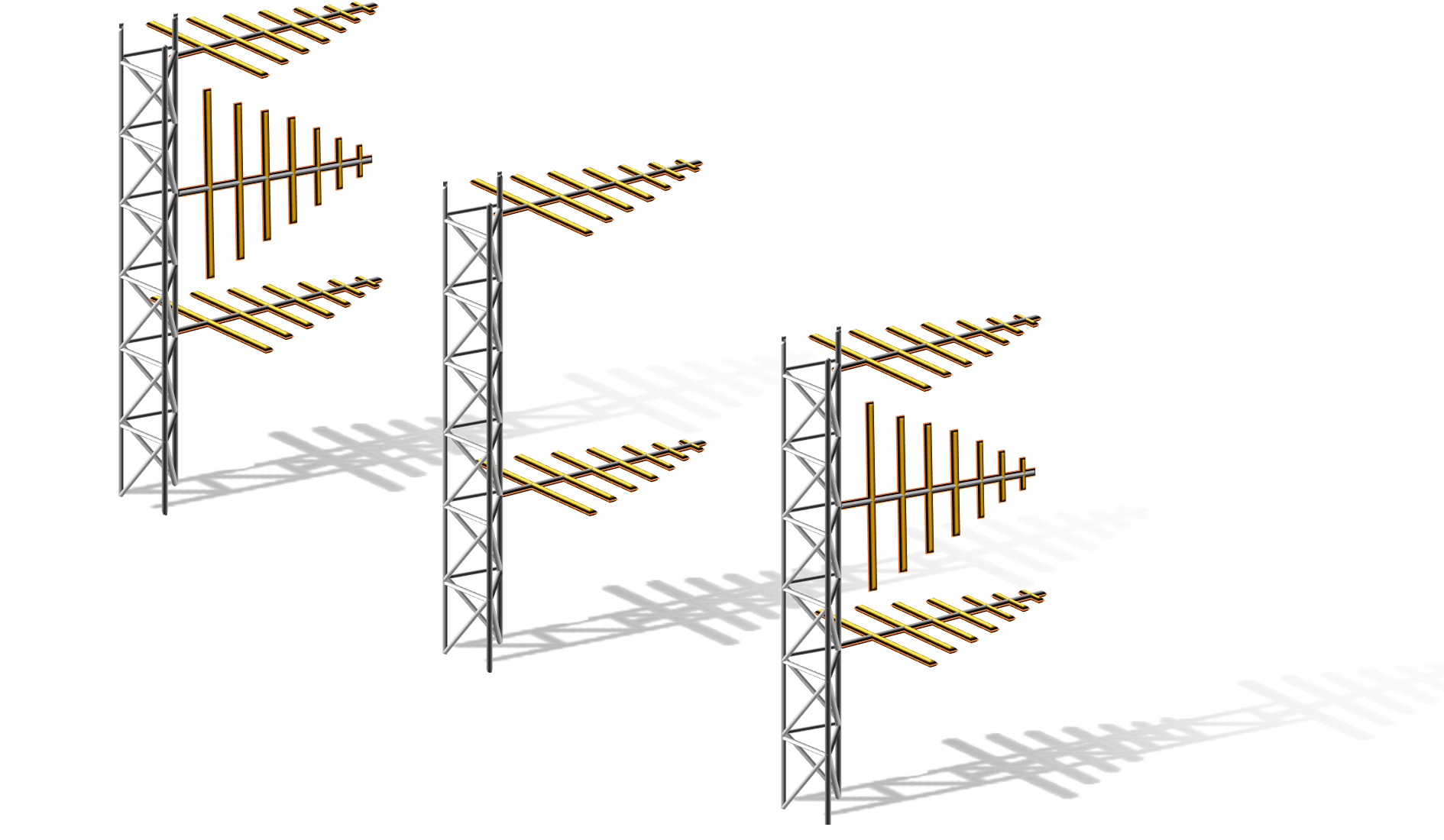}
    \includegraphics[width=0.4\textwidth]{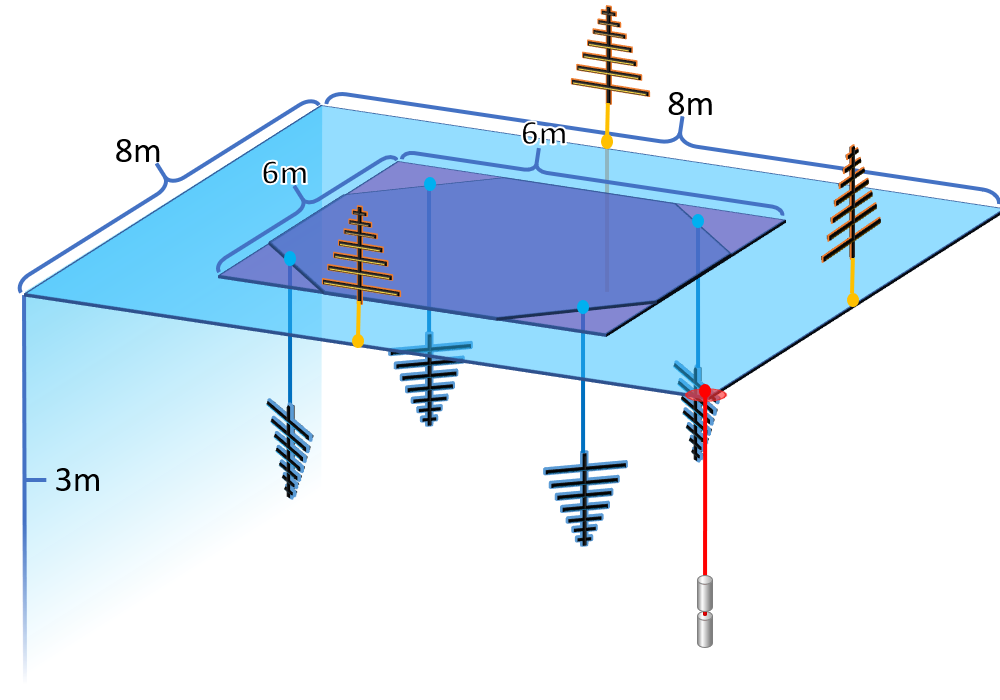}
    \caption{Overview of ARIANNA stations. (upper left) dedicated cosmic-ray station 32 at Moore's Bay, all antennas are in the ice (upper right) 8 channel neutrino/cosmic ray combo station 52 at Moore's Bay, all antennas are in the ice (lower left) horizontal cosmic-ray station 50, all antennas are above the ice (lower right) first ARIANNA station 51 at the South Pole, the downward facing antennas and the dipole are in the ice whereas the upward facing antennas are above the ice.}
    \label{fig:stations}
\end{figure}

In addition to the 7 HRA stations with 4 downward facing LDPAs that serve as a pilot array of a large scale neutrino detector, several other station designs have been installed for different science cases (cf. Fig.~\ref{fig:stations}).

The capabilities of cosmic-ray measurement were first demonstrated using a dedicated cosmic-ray station (station 32) consisting of four upward facing LPDAs \citep{Barwick2017} before the ARIANNA hardware was extended to support 8 channels. 
A year later in the 2017/2018 season, the first 8 channel station (station 52) was deployed combining 4 downward and 4 upward facing LPDAs.

In the same season, the first ARIANNA station was installed at the South Pole. Similar to station 52, it combines downward and upward facing LPDAs complemented by one dipole for a direct access of the vertical polarization. The upward facing LPDAs are placed above the surface to better study the RF noise situation at the South Pole. The station is connected to the power grid of the ARA detector so it can run all year. Since its deployment, the station has been running continuously proving that the ARIANNA hardware works reliably at the even colder temperatures at the South Pole which has average temperatures of \SI{-60}{\degree C} during the winter compared to \SI{-25}{\degree C} at Moore's Bay. The station was deployed relatively close to the South Pole station and in direct vicinity to a large wind turbine\footnote{This wind turbine is a kW scale power system from a different experiment and completely different from the low-power system that is discussed in Sec.~\ref{sec:windgen}.} that turned out to be a significant source of RFI noise. Although these events could be identified in an offline analysis it complicates data taking, limits the uptime in certain periods and results in larger data volumes. 

In the 2018/2019 season, another ARIANNA station was installed at the South Pole further away from any man made infrastructure. The station is now running completely autonomous with solar power and a battery to buffer periods of insufficient sunlight which results in a 100\% uptime during the austral summer. Also the second ARIANNA station has been running reliably since it was turned on. Being further away from South Pole station and the wind turbine has had the desired effect and we observe stable trigger rates. These preliminary results indicate that the South Pole is a suitable location for an ARIANNA style neutrino detector. 

\subsection{Prototype of a dedicated tau neutrino detector}
Another promising technique of a flavor sensitive neutrino detector \citep{Feng2002, HOU2002, Nam2017} is explored at the ARIANNA site similar to the approach of \citep{GrandWhitePaper2018}. A tau neutrino interaction produces a tau lepton which has just the right combination of interaction and decay length that it can escape a solid medium, then decay in air and produce an air shower that is in-turn observable with a radio detector. A solid mountain range provides an excellent target material. The Moore's Bay site provides an optimal location as it is surrounded by the massive Transantarctic Mountains and is extremely radio quiet. 

We prototype this technique with a horizontal cosmic-ray (HCR) station that consists of an array of antennas pointed towards the mountain range. Cosmic-ray induced air showers constitute the main physical background of such a tau neutrino detector. Hence, a precise measurement of air showers, in particular the angular reconstruction to distinguish air showers coming from slightly above the mountain from tau induced air showers coming from below the mountain ridge, is crucial to study the feasibility of this technique. 

In the 2017/2018 season, an earlier HCR prototype station \citep{Wang2017} was extended to the layout depicted in Fig.~\ref{fig:stations} bottom right. This station consists of eight {LPDAs} placed above the snow pointing at the mountain range that surrounds Moore’s Bay. An initial analysis of data from this improved station design indicates an improvement of the angular resolution in elevation from \SI{0.63}{\degree} to \SI{0.25}{\degree} for signals originating from close to the horizon. However, we observe a systematic offset of a couple of degrees which is mostly the result of interference with signals reflected off the surface and due to uncertainties in the station geometry. A full analysis of the HCR station will be presented in a forthcoming publication. 

\subsection{Wind power system}
\label{sec:windgen}
A wind power system would allow an autonomous station to run during the antarctic winter when the station is in darkness for almost 6 months. This has a huge potential as it directly doubles the uptime of the detector and its multi-messenger sensitivity. 
A wind generator capable of both surviving extreme Antarctic weather conditions and providing power at low wind-speeds has been in development since several years. 

\begin{figure}
    \centering
    \includegraphics[width=0.3\textwidth]{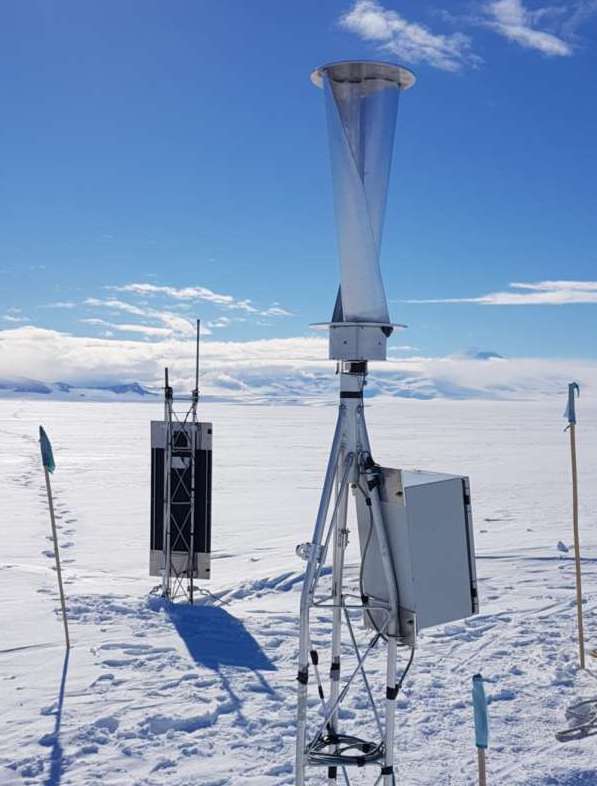}
    \includegraphics[width=0.69\textwidth]{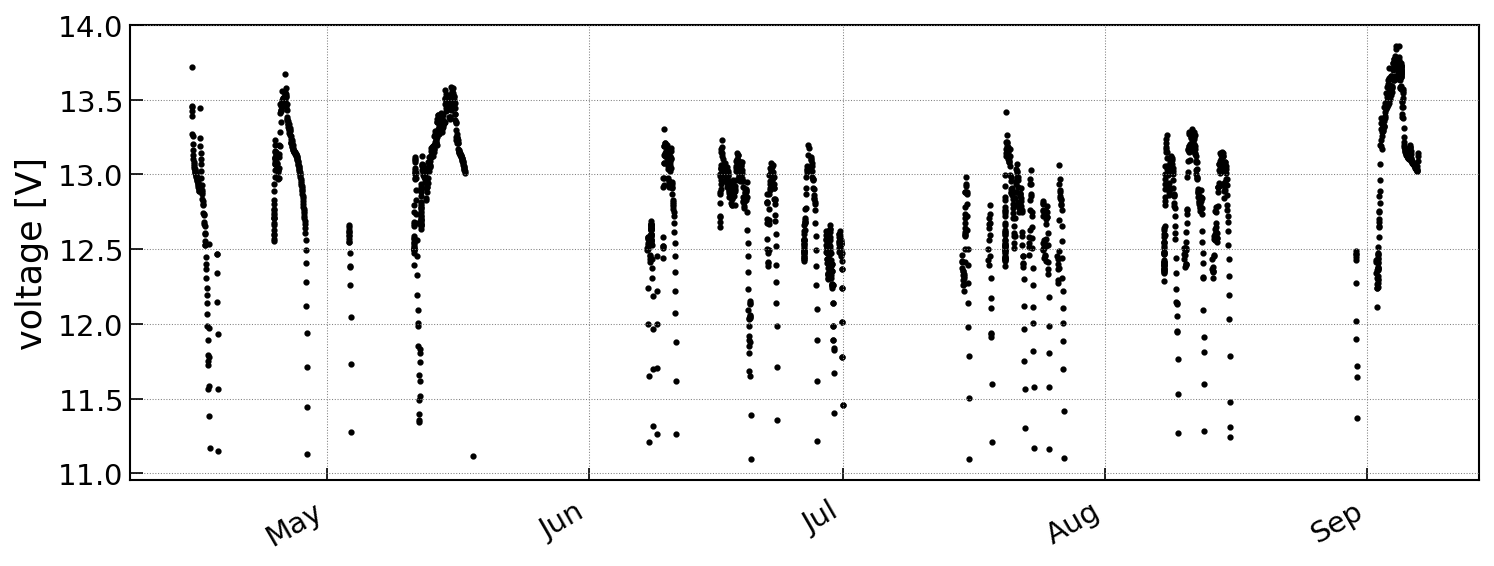}
    \caption{(left) Photo of newest version of the wind generator installed in Nov 2018 at Moore's Bay. (right) Shows operation of ARIANNA station throughout the dark winter (April-Sept) in 2018. During this part of the year, power was supplied entirely by Wind Generator.}
    \label{fig:windgen}
\end{figure}

The turbine has a novel geometry based on the traditional twisted Savonius
turbine. Its main elements are two twisted displaced half circle arc
cross-sections, in the horizontal plane, that are swept vertically with an
increasing azimuth around the axis of rotation (cf. Fig.~\ref{fig:windgen} left). This twist helps to reduce vibrations, noise
and the torque fluctuations. Above all it renders the turbine
aerodynamically self starting.
Integrated with the turbine is the permanent magnet rotor of an air gap
winding synchronous generator. This allows for a full electrical
control of the turbine's rotational speed through the loading of the generator.

The first important milestone was achieved in 2018 when the first wind generator survived the winter months and powered an ARIANNA station for 24\% of the winter time (cf. Fig.~\ref{fig:windgen} right).
In Nov 2018, a larger version was deployed at Moore’s Bay with several changes to improve performance at low temperatures which is shown in Fig.~\ref{fig:windgen} left. Its dimensions are \SI{0.6}{m} tall and \SI{0.3}{m} diameter with a weight of \SI{10.5}{kg}. We connected the new wind generator as the only power source to an ARIANNA station and found that the new system delivered sufficient power for 44\% of the time between Dec. 2018 and the time of writing this article of Jan. 2019, a time period of the year where windspeeds are generally lower than during the winter. Hence, a significant increase in uptime is expected during the winter.

Based on these encouraging results, we have designed a further upgrade that will be able to power an ARIANNA station for 70\% of the time at the South Pole (thus, an autonomous power system will provide power for 85\% of the year). The South Pole is more demanding because of lower temperatures, lower air density and slower windspeeds. Well understood scaling relations were used to estimate the required dimensions of \SI{1.2}{m} tall x \SI{0.6}{m} diameter and a weight of \SI{40}{kg} with estimated costs of \$5000. Hence, fully autonomous stations are a viable option for a large scale Askaryan detector at the South Pole.

\section{Reconstruction of signal direction}
\label{sec:direction}
In this section, we present a measurement campaign performed at the South Pole during the austral summer in 2017/2018. This campaign measured the propagation of radio signals from deep in the ice to the surface and constitutes an important proof-of-concept of a surface array. Furthermore, it provides important insights into the understanding and modelling of signal propagation through the ice and in particular the firn. In the upper ice layers (the firn), the index of refraction changes from $n = 1.78$ of deep ice to $n = 1.35$ at the surface which results in a bending of the signal trajectories due to continuous Fresnel refraction. This effect is modelled using ray optics. Both the modeling via ray optics as well as the description of the index-of-refraction profile can be tested using this measurement. 

The measurement is set up as follows: A pulser, i.e., an antenna that emits short broadband pulses, was lowered in a hole \SI{653}{m} away from station 51. The cored hole drilled by the SPice experiment \citep{SPice2014} is \SI{1.75}{km} deep which allowed to place the pulser deep enough to be outside of the \emph{shadow zone}. Signal trajectories are bent downwards due to the changing index of refraction, such that signals emitted from a shallow depth are not able to reach the ARIANNA station. We note that exceptions from this classical picture have been observed (see e.g. \cite{Barwick2018}) and indeed also signals emitted from within the shadow zone are visible in this measurement. However, an analysis of this data is beyond the scope of this article and will be addressed in an upcoming publication. Here, we focus on the 'classical' region that is important for the default operation of a radio neutrino detector. 

This data is used to reconstruct the signal arrival direction which is then compared to the predicted directions calculated from the known geometry and ray tracing the signal though the ice. 
We analyzed 51 pulses emitted from a depth between \SI{761}{m} and \SI{841}{m}. The signal arrival direction was reconstructed from the time differences of the two pairs of parallel channels of the downward pointing LPDAs. The time differences were determined using a cross-correlation method. The advantage of this method is that it is independent of the description of the antenna response as only the time differences of parallel channels are considered where the antenna response is the same and thus cancels out as systematic uncertainty. Please refer to \citep{PersichilliPhD} for more details on the analysis and \citep{NuRadioReco} for details of the reconstruction algorithm.

\begin{figure}[t]
    \centering
    \includegraphics[width=0.3\textwidth]{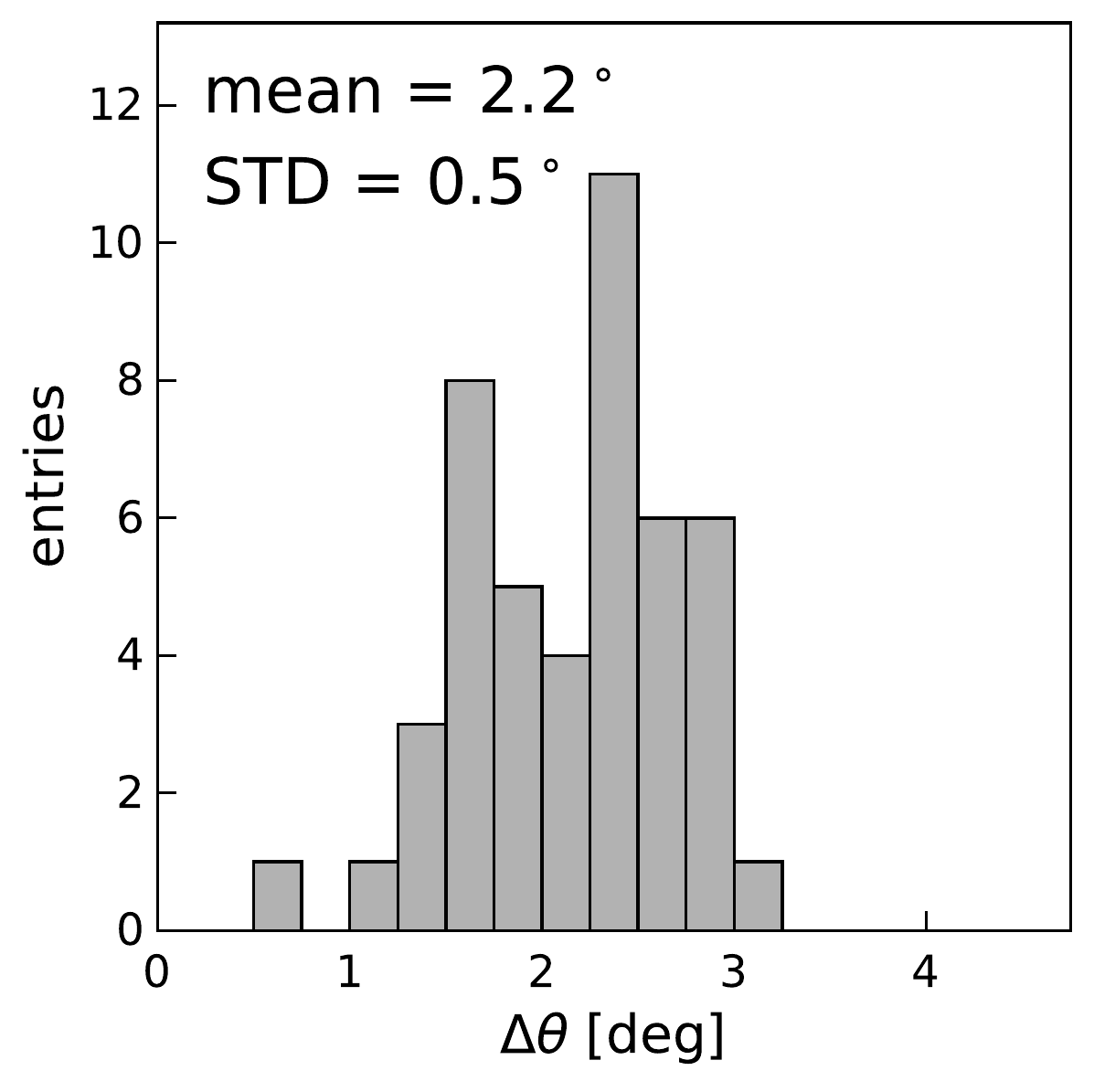}
    \includegraphics[width=0.3\textwidth]{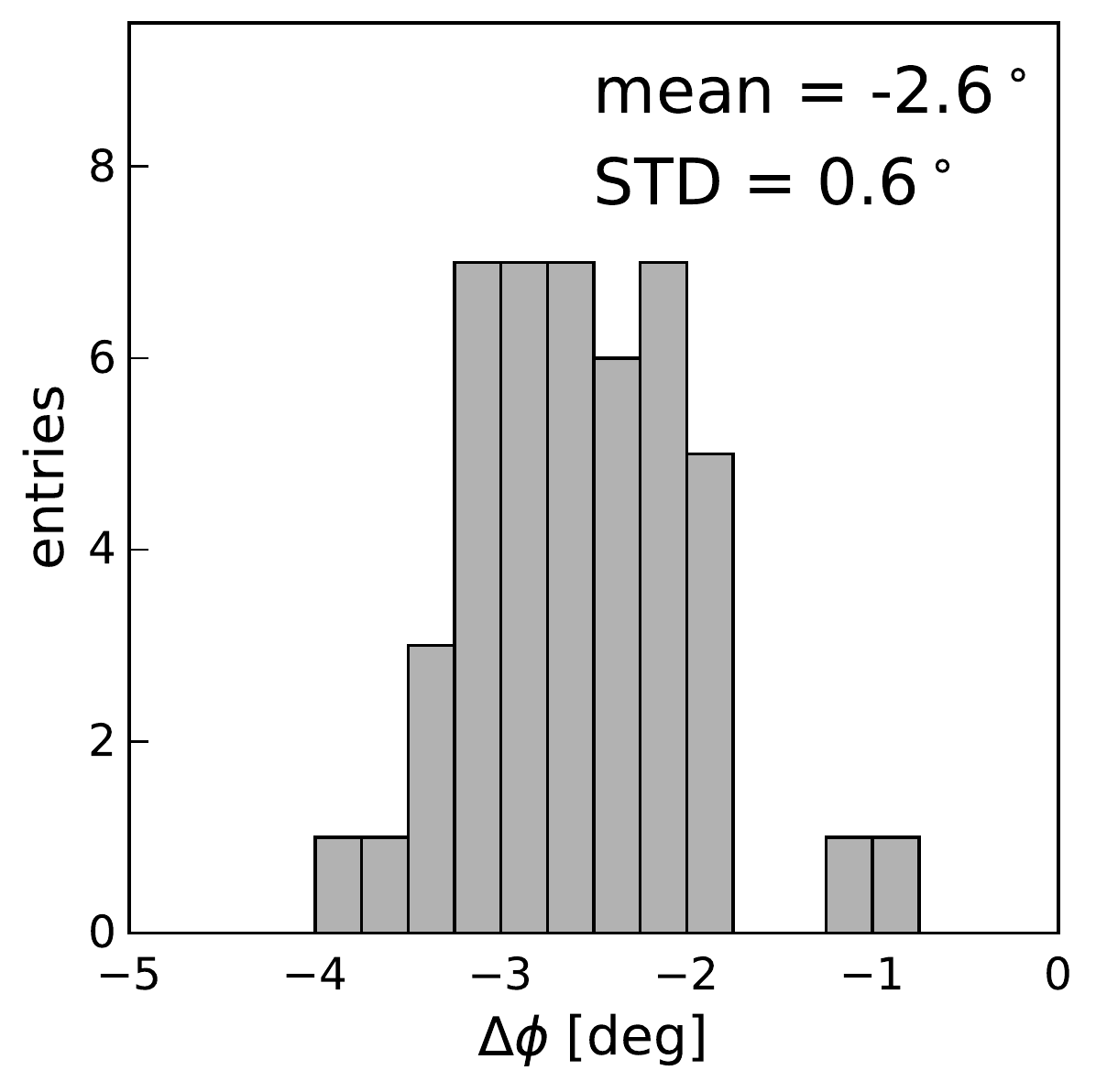}
    \caption{Experimentally determined angular resolution from an in-situ measurement at the South Pole. The histograms show the difference between the expected and reconstructed direction of the zenith angle (left) and azimuth angle (right).}
    \label{fig:angularres}
\end{figure}

For each pulse, we determine the expected arrival direction via ray tracing the signal through the ice, and compare it with the reconstructed direction. The result is presented in Fig.~\ref{fig:angularres} and we find a scatter of less than \SI{1}{\degree}. This has been achieved despite the small lever arm of only \SI{6}{m} because of the excellent time synchronization across channels of the ARIANNA hardware of better than \SI{5}{ps}. 

We observe an offset of \SI{2.6}{\degree} in the azimuth angle and \SI{2.2}{\degree} in zenith angle.
We thoroughly evaluated all experimental uncertainties and can't exclude that the observed offsets are due to uncertainties in the detector calibration and position of the emitter. However, the offset in the zenith angle might also be due to uncertainties in the calculation of the expected signal arrival direction, e.g., from uncertainties in the assumed index of refraction profile. From the experience gained in this calibration campaign, we can improve the detector calibration in the future such that it won't limit the uncertainties in the directional reconstruction.

The main conclusions with relevance to a radio neutrino detector from this measurement are: First, radio signals propagate from deep in the ice to the surface, i.e., neutrinos can be observed via an array of antennas placed just slightly below the ice surface. 
Second, the index-of-refraction profile is understood well enough and our mathematical modelling of the signal propagation via ray optics is accurate enough to predict the incoming signal direction within two to three degrees and might even be much better as some of the offset can originate from systematic uncertainties in the detector calibration or position of the emitter. 

This result underlines the importance to precisely measure the ice properties and to understand the propagation of radio signals through the ice. 
During the last austral summer of 2018/19, an improved measurement was performed that built up on the experiences gained with the measurement presented here. This new measurement will allow us to study propagation effects on the pulse form and the study of birefringence effects. We will report on this new measurement in an upcoming publication. 

We note that the signal arrival direction is different from the neutrino direction because the radio signal is emitted on a cone around the neutrino direction with an opening angle of about \SI{56}{\degree}. To determine the neutrino direction, we also need to know the polarization of the radio signal, which is addressed in Sec.~\ref{sec:CRs}. Given the angular direction of the arriving signal and the polarization, the neutrino direction can be determined.  It is limited by the few degree width of the Cherenkov cone, though it may be possible to constrain the angular offset from the axis of the Cherenkov cone by examining the frequency dependence of Askaryan pulse.

\section{Search for neutrino signals}
\label{sec:template}
\begin{figure}
    \centering
    \includegraphics[width=0.4\textwidth]{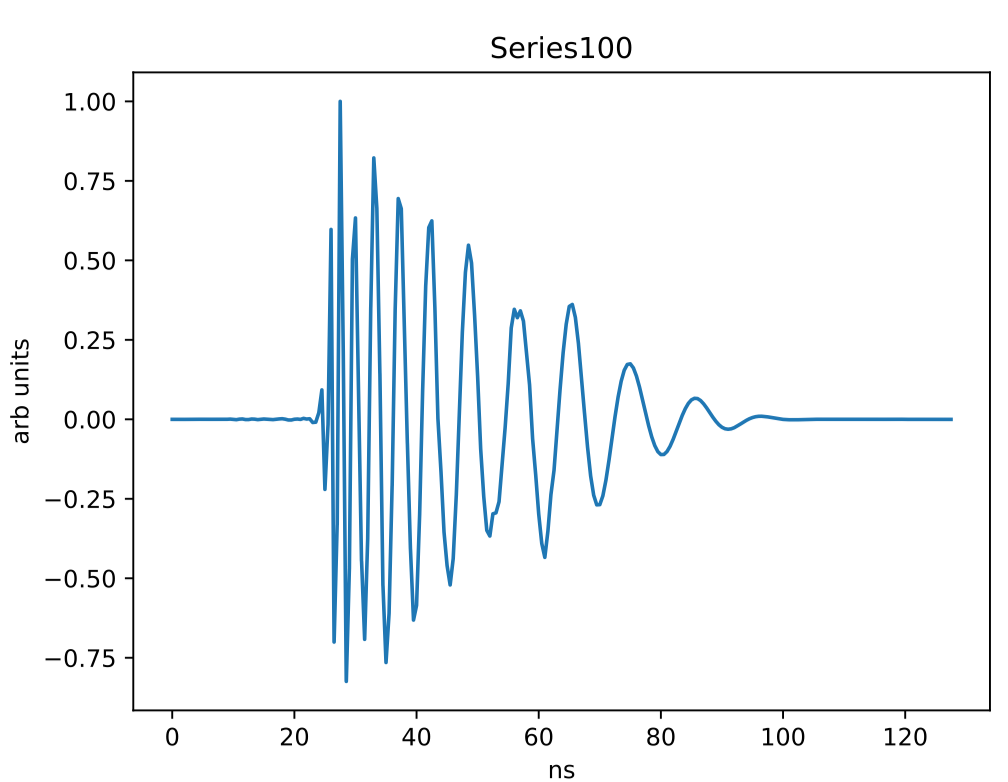}
    \caption{ARIANNA detector response to a on-cone neutrino radio pulse (cf. Fig.~\ref{fig:askaryan}). The LPDA antenna response was evaluated \SI{30}{\degree} off the boresight direction which corresponds to the most likely signal arrival direction. The high frequency components are at the beginning of the waveform whereas lower frequency components are delayed longer and are shifted towards the end of the waveform. Figure from \citep{PersichilliPhD}.}
    \label{fig:template}
\end{figure}

The unambiguous identification of radio pulses originating from a neutrino interaction in the ice is the first purpose of the ARIANNA detector. Already the distinctive detector response to an impulsive signal allows for an efficient discrimination of neutrino candidates against the thermal and anthropogenic noise background. In particular, the dispersion of the LPDA antenna, which comes as an unavoidable side effect of the LPDA's high gain, produces a waveform that is clearly distinguishable from thermal noise fluctuations and from most anthropogenic radio pulses. The ARIANNA detector response to an Askaryan pulse, which we refer to as \emph{neutrino template} in the following, is presented in Fig.~\ref{fig:template}.

The similarity of a triggered event with the neutrino template is determined by means of the Pearson correlation coefficient $\chi$ which can take values from 0 to 1, where 1 represents perfect correlation and values close to 0 represent totally un-correlated waveforms.

\begin{figure}
    \centering
    \includegraphics[width=0.49\textwidth]{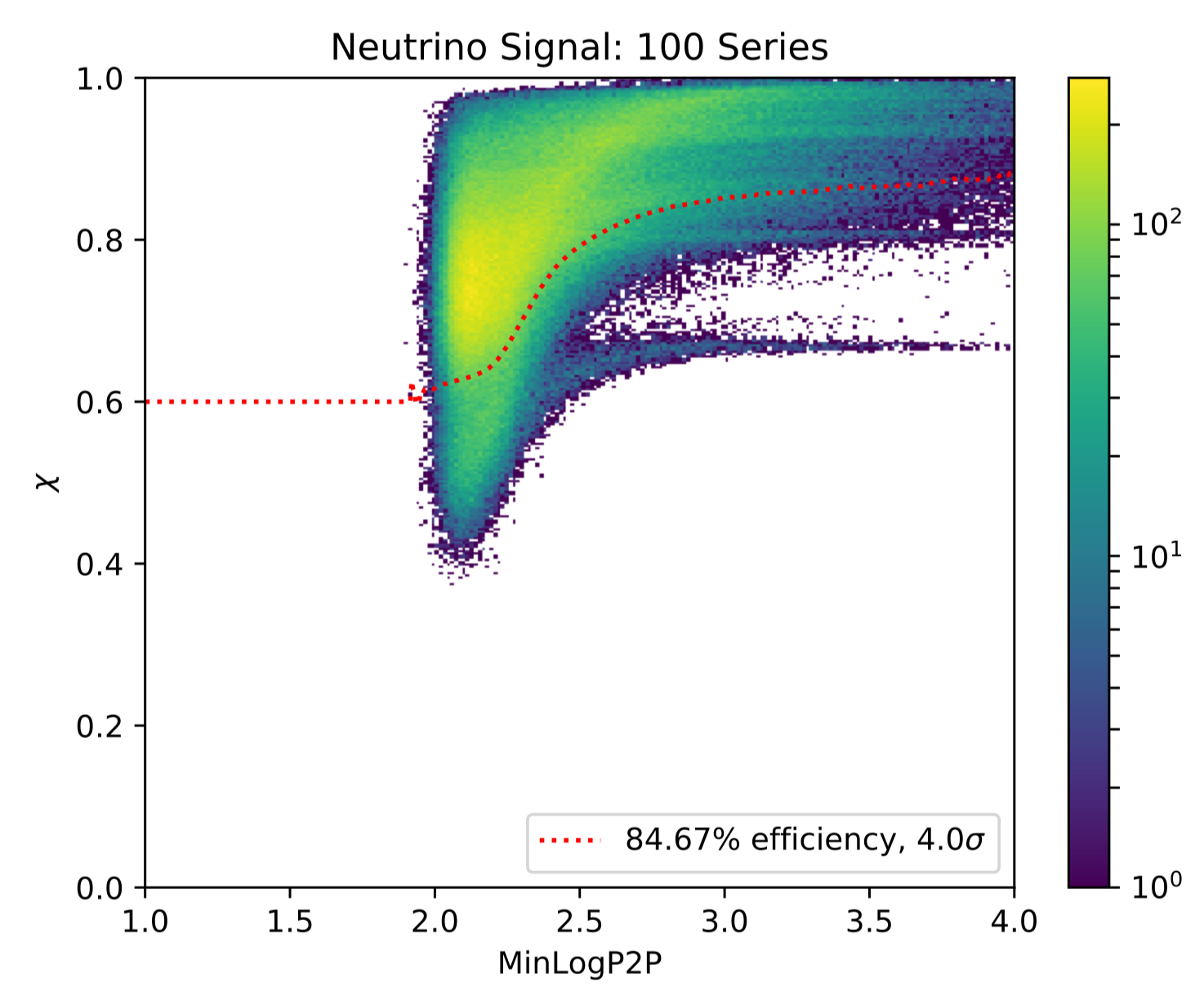}
    \includegraphics[width=0.49\textwidth]{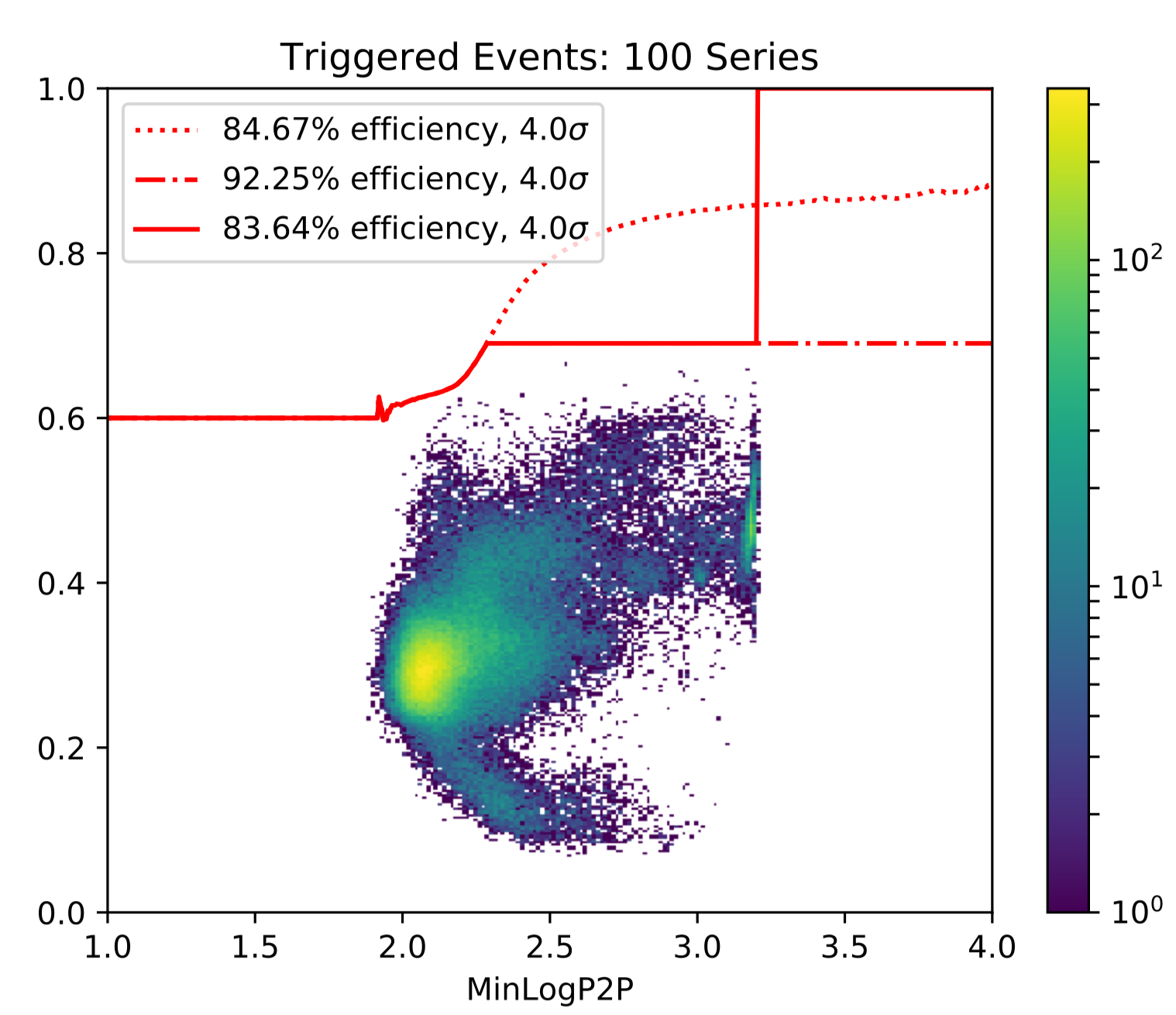}
    \caption{(left) Distribution of expected neutrino Askaryan signals in the similarity parameter $\chi$ and peak-to-peak amplitude. The color shows the number of entries per bin. The area above the red curve contains 85\% of the simulated neutrino signals. (right) Same as the left plot but for triggered events. Figure from \citep{PersichilliPhD}.}
    \label{fig:chidist}
\end{figure}

The similarity estimator $\chi$ is combined with an estimator of the signal strength, because high amplitude signals are less effected by noise and are therefore expected to have a higher $\chi$ coefficient than a neutrino pulse that hardly sticks out of the noise. Hence, for signals with a high signal-to-noise ratio, we can require a higher $\chi$ value. The optimal cut value is determined in a separate Monte-Carlo study in which a detector simulation is performed for a large library of simulated Askaryan signals. The resulting distribution of $\chi$ vs. the peak-to-peak amplitude (P2P) is shown in Fig.~\ref{fig:chidist} left.

This neutrino signal distribution is compared to the distribution of all triggered events. We analyzed all data from Dec. 2015 through April 2017. The corresponding distribution from all HRA stations equipped with a series 100 amplifier\footnote{The ARIANNA HRA stations have two different types of amplifiers with a slightly different gain which are referred to as series 100 and 200 amplifiers. Newer stations were deployed with the series 200 amplifiers.} is shown in Fig.~\ref{fig:chidist} right. The distribution of the stations equipped with a series 200 amplifier look qualitatively the same. The bulk of the events is clearly separated and far away from the expected neutrino signal space, and none of the recorded events reaches the approx. 85\% neutrino efficiency line. This analysis demonstrates that already a simple template matching technique is a powerful discriminator that leads to a good neutrino efficiency and purity. 

However, an important physical background to the template matching technique is the radio emission of cosmic-ray air showers that is picked up by the in ice antennas. Their radio pulse is very similar to the expected Askaryan signal and cosmic rays are several orders of magnitude more abundant than neutrinos with a typical event rate of one per day per ARIANNA station. The initial cosmic-ray signal always comes from above and enters the (downward facing) LPDA antennas through their backlobe. This normally distorts the signal strongly enough to be distinguishable from Askaryan signals that come from below and enter the antenna through its sensitive direction. However, for some very specific air-shower directions, the cosmic-ray pulse might be confused with a neutrino signal, and a preliminary analysis actually observed one event that is in the neutrino signal space. Therefore, the newer generation of ARIANNA stations is equipped with four additional upward facing LPDAs which allows a clear tagging of cosmic-ray events. A simple cut on the amplitude ratio between upward and downward facing antennas reduces the cosmic-ray background to 0.3 neutrino candidates in 3 calendar years for an array with more than 1000 detector stations while preserving 99.7\% of the neutrino triggers \citep{ARIANNA2014}.

\section{Cosmic ray test beam\footnote{For completeness of this article, this section partly reviews previously published results on the detection and selection of cosmic rays of \citep{Barwick2017}.}}
\label{sec:CRs}
The measurement of cosmic rays comes with many advantages apart from rejecting cosmic-ray signals from the neutrino search. Cosmic rays are not only a background that we need to get rid of but also a perfect calibration source for a radio-neutrino detector because their radio pulses are very similar to the Askaryan pulses that we expect from neutrinos. Both are very short bipolar pulses of just a few nanoseconds length which are difficult to generate artiﬁcially. Therefore, measuring cosmic rays is the only way to fully test the neutrino detector under realistic conditions. Furthermore, the radio emission of air showers is well understood so that the reconstructed signal properties can be verified by theoretical predictions.

\begin{figure}
    \centering
    \includegraphics[width=0.49\textwidth]{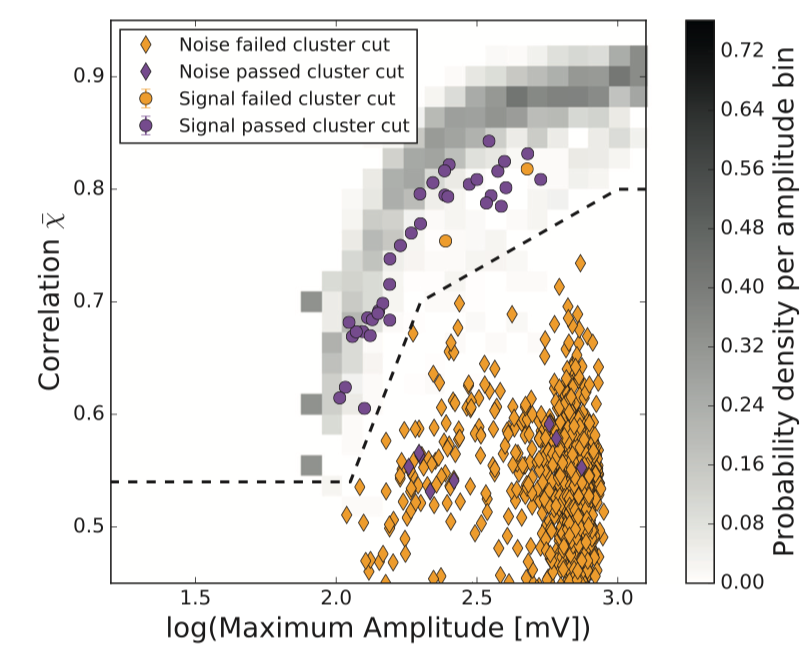}
    \includegraphics[width=0.49\textwidth]{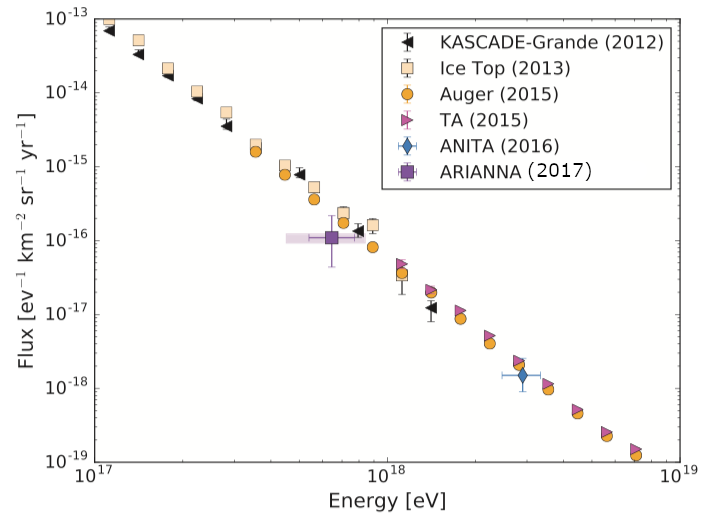}
    \caption{(left) Correlation of measured events with a cosmic-ray template as a function of signal amplitude. The background density map shows the probability distributions for simulated air showers given an amplitude. The markers show the average correlation value $\chi$ of measured events. The line indicates a cut separating the cosmic ray signals from the background. All diamonds are background events, while the signals are indicated by circles. See \citep{Barwick2017} for more details. (right) Cosmic-ray flux measured by ARIANNA in comparison with other experiments. Figures and captions from \citep{Barwick2017}.}
    \label{fig:CRs}
\end{figure}

First, we demonstrate the performance of the template matching technique by using this technique to identify cosmic-ray signals out of the large sample of all triggered events \citep{Barwick2017}. A two dimensional cut in the correlation parameter $\chi$ and the signal amplitude leads to a clear separation of cosmic rays from the background as shown in Fig.~\ref{fig:CRs} left. The measured event rate together with a simulation of the ARIANNA acceptance and the uptime of the detector was converted to a cosmic-ray flux and shown in Fig.~\ref{fig:CRs} right. The measured flux agrees with the more precise measurements of other experiments within uncertainties which is an indirect test that the cosmic-ray identification works successfully.

Second, we demonstrate the ARIANNA sensitivity to the signal polarization by reconstructing the polarization of the cosmic-ray events and comparing it with the theoretical expectation \citep{GlaserARENA2018}. The polarization of cosmic-ray radio signals was measured extensively by dedicated radio cosmic-ray detectors such as AERA \citep{AERAPolarization} and LOFAR \citep{LofarPolarization2014} and is very well understood theoretically \citep{Vries2010b, GlaserErad2016, Glaser_2019} which allows for a precise prediction.

We use all data collected by the dedicated cosmic-ray station 32 during the 2017/2018 season and select cosmic rays using the template-matching method described above. In total we find 265 cosmic-ray events from which 135 pass the more stringent quality cut of having a signal-to-noise ratio (SNR)\footnote{The signal-to-noise ratio is defined as the maximum amplitude divided by the RMS noise.} larger than 4 in all four channels. (We note that a SNR of 4 is a relatively weak cut, the average SNR of a pure noise trace is 3.) This cut allows to reconstruct the cosmic-ray direction using the same method as described in Sec.~\ref{sec:direction}. The direction is required to predict the expected signal polarization, and to evaluate the antenna response for the correct direction during the polarization reconstruction. 

We reconstruct the signal polarization using a novel forward folding method that is described in detail in \citep{NuRadioReco} and briefly summarized here: The incident electric field, i.e., the cosmic-ray radio pulse, is related to the measured voltage in the antennas in Fourier space via 
\begin{equation}
    \begin{pmatrix} \mathcal{V}_1(f) \\ \mathcal{V}_2(f) \\ ...\\ \mathcal{V}_n(f)\end{pmatrix} = 
    \begin{pmatrix} \mathcal{H}_1^\theta (f)& \mathcal{H}_1^\phi (f)\\ \mathcal{H}_2^\theta (f) & \mathcal{H}_2^\phi (f)\\ ... \\ \mathcal{H}_n^\theta (f)& \mathcal{H}_n^\phi (f)\end{pmatrix} 
    \begin{pmatrix} \mathcal{E}^\theta(f) \\ \mathcal{E}^\phi(f)\end{pmatrix} \, ,
    \label{eq:H_full}
\end{equation}
where $\mathcal{V}_i$ is the Fourier transform of the measured voltage trace of antenna $i$, $\mathcal{H}_i^{\theta, \phi}$ represents the response of antenna $i$ to the $\phi$ and $\theta$ polarization of the electric field $\mathcal{E}^{\theta, \phi}$ from the direction $(\varphi,\vartheta)$.

Traditionally, this system-of-equation is solved for the electric field for each frequency bin. However, this unfolding comes with several downsides and often leads to an amplification of noise, in particular, for horizontal air showers and higher frequency bands where the cosmic-ray signal is small (see extended discussion in \citep{NuRadioReco} for more details). Instead, we use a novel forward folding technique: We describe the cosmic-ray radio pulse analytically with just four free parameters, and determine these parameters in a chi-square minimization directly on the measured voltage traces. This is, the analytic electric-field is folded with the antenna response to obtain the expected signal in all antennas. These voltage traces are then directly compared to the measured voltage traces. The parameters of the electric-field pulse are optimized to obtain the smallest quadratic difference (the $\chi^2$) between the two traces for all channels. 

The polarization angle is then just given by the ratio of the two electric-field components. Applying this method on the cosmic-ray data set of station 32 and comparing it to the theoretical expectation, we find a resolution of the signal polarization of \SI{14}{\degree}. The expected polarization is calculated according to the dominant geomagnetic emission process and is given by the vector product between the shower axis and the geomagnetic field axis. We expect the resolution to improve significantly with an improved detector calibration. To study this, we performed an end-to-end MC simulation using a representative set of CoREAS simulations \citep{Coreas} and including signal distortion due to noise interference. For a well-calibrated detector station we find that we can achieve a polarization resolution of $\sim3^\circ$ which is limited by the signal-to-noise ratio.

Although not the primary objective of ARIANNA, a direct contribution to ultra-high-energy cosmic-ray (UHECR) physics is foreseen. A large-scale ARIANNA detector with hundreds of stations will provide a substantial exposure to measure cosmic rays with reasonable statistics up to energies of \SI{e19}{eV} \citep{Barwick2017}. We will be able to measure the energy spectrum with competitive and independent systematic uncertainties \citep{Gottowik2017}. Furthermore, a small subset of cosmic rays are expected to be observed in coincidence with IceTop and IceCube if the detector is deployed at the South Pole. These events will provide an exceptional amount of information and will be important for cross-calibration purposes. In addition, the radio signal is only sensitive to the electromagnetic air-shower component whereas IceTop and especially IceCube provides a complementary measurement of the hadronic shower component which allows for a measurement of the cosmic-ray mass \citep{Holt2019}. This is facilitated by a newly developed technique to determine the air-shower energy from a single radio detector station \citep{Welling2019}. Also, the radio signal of inclined air showers, where this technique works best, extends over a large area \citep{AERAHorizontal2018}. Thus, a sparse spacing of detector stations of $\mathcal{O}$(\SI{1}{km}) is sufficient to perform this measurement.

\section{ARIA: Optimized surface radio neutrino detector}
The versatile experiences of the ARIANNA pilot array in terms of hardware stability, deployment and detector operation in the harsh Antarctic conditions, neutrino and cosmic-ray identification, and advanced data reconstruction led to the design of the Askaryan Radio In-ice Array (ARIA). We determined the South Pole as the optimal location because of the good infrastructure and the cold, deep and clear ice which attenuates the radio signal less than the ice at the Moore's Bay site and more than compensates for the higher logistical effort and lack of reflective layer from the water-ice interface beneath the ice shelf at Moore's Bay. 

\begin{figure}
    \centering
    \includegraphics[width=0.6\textwidth]{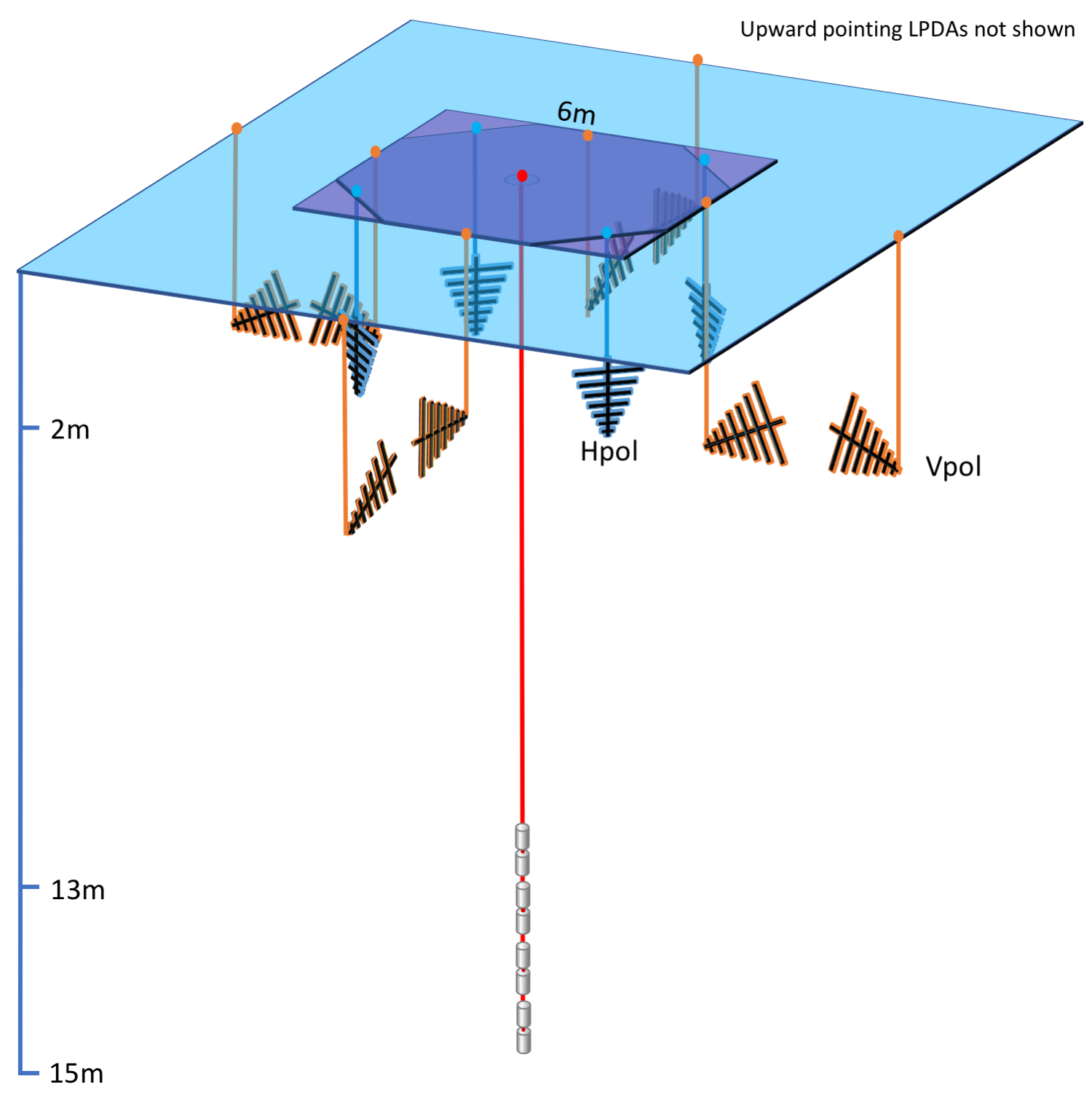}
    \caption{Sketch of the ARIA station design.}
    \label{fig:ARIA}
\end{figure}

ARIA combines the advantages of the ARIANNA design (autonomous stations, ease of deployment, reliable hardware, deployable at any ice sheet on the planet) with an optimized station layout that achieves a high sensitivity to neutrinos, an unambiguous identification of the neutrino signals with multiple complementary channels, and the ability to reconstruct the neutrino direction and energy for almost all events. Because no high-energy neutrino has been detected yet with the radio technique, especially the second point is of utmost importance. All these goals can be achieved with the station design presented in Fig.~\ref{fig:ARIA}. 

The ARIA station comprises several LPDA antenna with different orientations near the surface and one string of four dipoles deployed at a depth of \SI{15}{m}. The vertically oriented LPDAs provide a large sensitivity to the expected neutrino signals as the antenna is most sensitive towards the expected signal direction and polarization. The four downward pointing LPDAs measure the two horizontal polarization components and provide the necessary addition information to reconstruct the polarization. The four upward pointing LPDAs take care of cosmic-ray rejection. The station is then completed with four deep dipoles to improve the neutrino identification and provide information on the distance to the neutrino interaction vertex via detecting both the direct and reflected pulse (see below). 

The autonomous nature of the stations and the vicinity of the antennas to the snow surface allow for a quick and easy deployment with essentially no requirements for supporting infrastructure. The use of a newly developed portable cylindrical hole melter allows to drill the borehole for the dipoles within a few hours and very little monitoring \citep{Heinen2017, MeltingProbe}. It was already successfully used in 2018 at the Moore's Bay site. This concept is currently adapted to also melt the slots for the LPDAs to avoid hand digging of trenches. A complete ARIA station can be deployed within one working day with about four to five people. 

Neutrinos can be identified using several complementary techniques:
\begin{itemize}
    \item Using a template matching technique as described in Sec.~\ref{sec:template} where cosmic-ray signals can be separately rejected by the upward pointing LPDAs.
    \item The additional information of the signal arrival times from the deep dipoles allow to determine if the signal originated from above, i.e., cosmic rays or anthropogenic background, or from below where the neutrino signals come from.
    \item The deep dipoles measure two pulses, a direct and reflect pulse, which is unique signature of a neutrino signal (D'n'R technique, see below).
\end{itemize}

\subsection{D'n'R technique}
The direct and reflected (D'n'R) technique provides a unique signature of a neutrino signal and provides important information on the distance to the neutrino interaction vertex: For most geometries, signals originating from deep in the ice have two paths to an antenna. One direct path and one reflected path, i.e., the signal gets reflected off the surface and reaches the antenna from above. We studied this technique in detail using the novel Monte Carlo code NuRadioMC \citep{NuRadioMC} that models the signal propagation through the firn via ray tracing and includes a realistic treatment of the reflection at the surface using the complex Fresnel reflection coefficients. We find that for most neutrino events, the incident angle at the surface is such that is results in total internal reflection. Hence, the ice-air interface acts as a perfect mirror. 

The optimal depth of the dipoles is a trade off between two competing effects. On the one hand, the time difference between both pulses increases with depth which allows for a better differentiation and improves the resolution of the vertex distance. On the other hand, the efficiency to detect both pulses reduces with increasing depth. This is because of the narrowness of the Cherenkov cone. The optimal depth is around \SI{15}{m} where the ARIA dipoles are located. At this depth we get time differences between both pulses of a few tens of nanoseconds which allows for a clear separation. Even at low neutrino energies of \SI{e17}{eV}, the D'n'R detection efficiency is still above 80\%. At \SI{e18}{eV} the detection efficiency of both pulses is 95\% and even larger at higher neutrino energies. 

\begin{figure}
    \centering
    \includegraphics[width=0.45\textwidth]{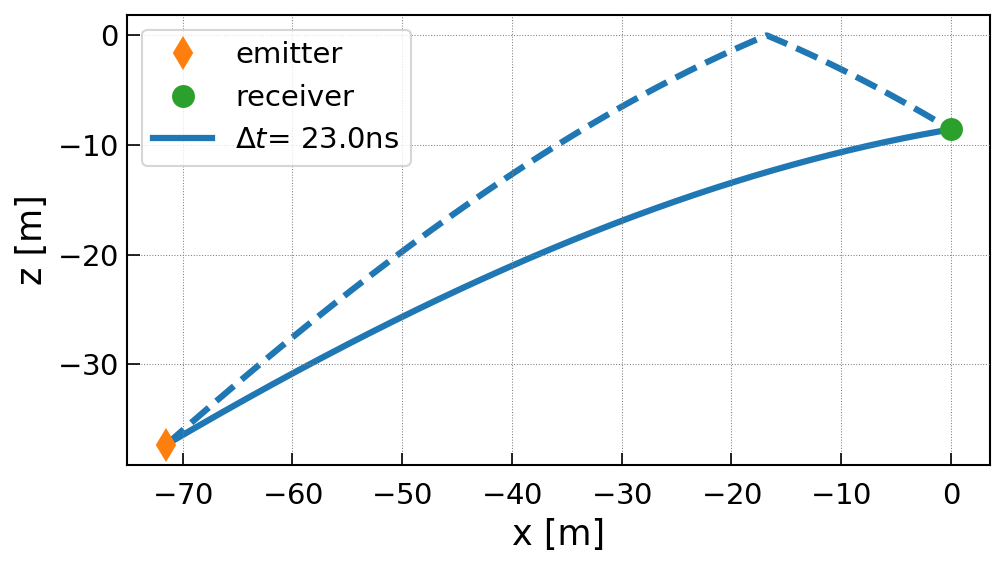}
    \includegraphics[width=0.54\textwidth]{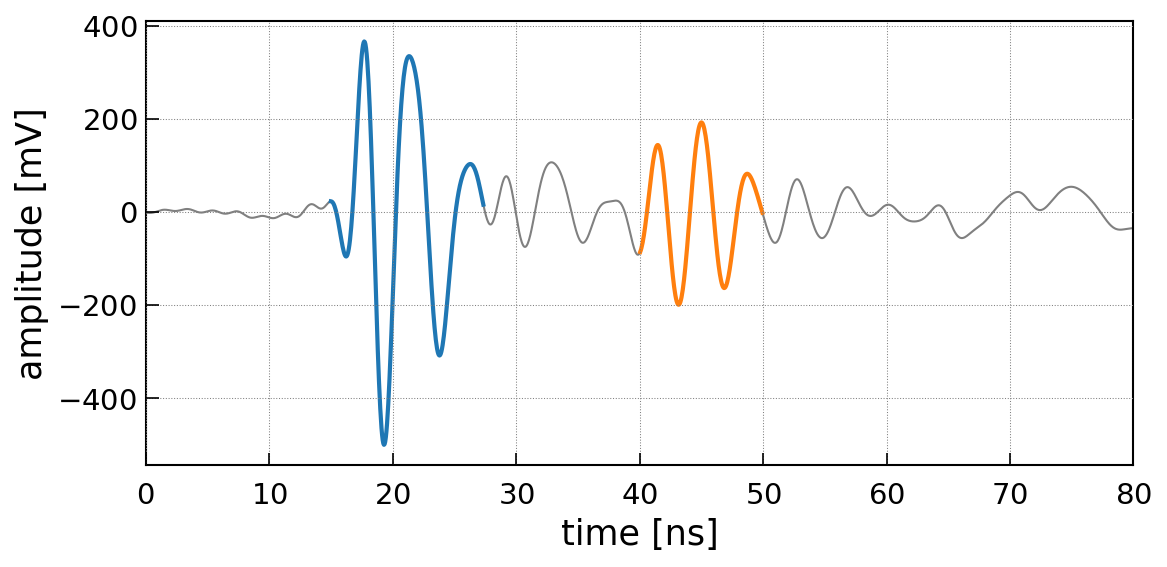}
    \caption{D'n'R pulser measurement at Moore's Bay. (left) Geometry of the setup. The direct path (solid line) and reflected path (dashed line) are calculated using the ray-tracing technique. (right) The received waveform in the \SI{8.9}{m} deep bicone antenna. Both the direct (blue) and reflected pulse (orange) are clearly observed.}
    \label{fig:DnR}
\end{figure}

Experimental evidence exists that reflected pulses can be observed in addition to direct pulses. The ARA collaboration broadcasted radio signals from a \SI{1400}{m} deep emitter to their receiving antennas at \SI{200}{m} depth and observed both pulses \citep{Abdul2017}. The measured time difference is compatible with the propagation times of the two paths that were determined via ray tracing. We note that this measurement implies that the signal propagated twice through the complete firn, and that the bending of signal trajectories in the firn is not hampering this technique. 
More recently, we performed a similar measurement at the Moore's Bay site where a short broadband pulse was emitted at a depth of \SI{37.5}{m} to a \SI{8.9}{m} deep receiving antenna. The geometry and the received signal is presented in Fig.~\ref{fig:DnR}. The two signal paths between emitter and receiver are calculated with ray tracing using the \emph{MB \#1} exponential index-of-refraction profile that describes the available index-of-refraction data well \citep{Barwick2017}. Technically, we used the analytic ray-tracing code of NuRadioMC \citep{NuRadioMC}. We find that already at \SI{8.9}{m} depth, both pulses are clearly observed and that the time difference between the two matches well the expected time delay of \SI{23}{ns} calculated via ray tracing.

\subsection{Reconstruction capabilities}
To determine the neutrino direction, both the signal arrival direction as well as the polarization need to be measured. The signal arrival direction is reconstructed from the signal arrival times of all antennas (cf. Sec.~\ref{sec:direction}). 
The polarization is reconstructed from the LPDA measurements using the same method as for cosmic rays (cf. Sec.~\ref{sec:CRs}). It is important to point out that the systematic uncertainty of this measurement will be small because the same antenna type is used to cover all three polarizations. The LPDA antennas are just oriented into different directions to be sensitive to different signal polarizations. We also note that the forward folding technique can naturally use all available antennas. Hence, we expect an improvement of the polarization resolution due to the additional antennas of an ARIA station compared to results of the previous section obtained with only four antennas. However, the method still needs to be adapted to neutrino signals that have a different pulse form. In addition, the pulse forms are slightly different between antennas due to the narrowness of the Cherenkov cone. This will be addressed in an upcoming publication. 

The determination of the neutrino energy is the most challenging part and requires the measurement of the distance to the neutrino interaction vertex, the viewing angle, i.e., which part of the Cherenkov cone is observed, and the signal polarization. The vertex distance is determined from the time difference of the direct and reflected pulse observed in the dipoles. The viewing angle is reconstructed from the frequency spectrum of the Askaryan signal. Here, the broadband sensitivity of the LPDA antennas is beneficial. Hence, the ARIA design has a good sensitivity\footnote{The energy resolution is physically limited by the random fluctuations of how much energy is transferred from the UHE neutrino into the particle shower that produces the Askaryan signal. This limits the energy resolution to about $\sigma[\log_{10}(E_\mathrm{rec}/E_\mathrm{true})] \sim 0.35$.  A preliminary analysis indicates that the contributions to the energy resolution from the uncertainty in vertex distance, viewing angle, and polarization are smaller than the aforementioned limit.} to the neutrino energy.

It is important to point out that ARIA is sensitive to the neutrino properties of essentially all triggered events. Each neutrino signal will be seen in three orthogonal LPDA orientations, which allows for a measurement of the polarization and frequency spectrum, and most events will have a direct and reflect pulse in the dipoles, which allows for the reconstruction of the vertex distance.

\subsection{Array layout}
\begin{figure}
    \centering
    \includegraphics{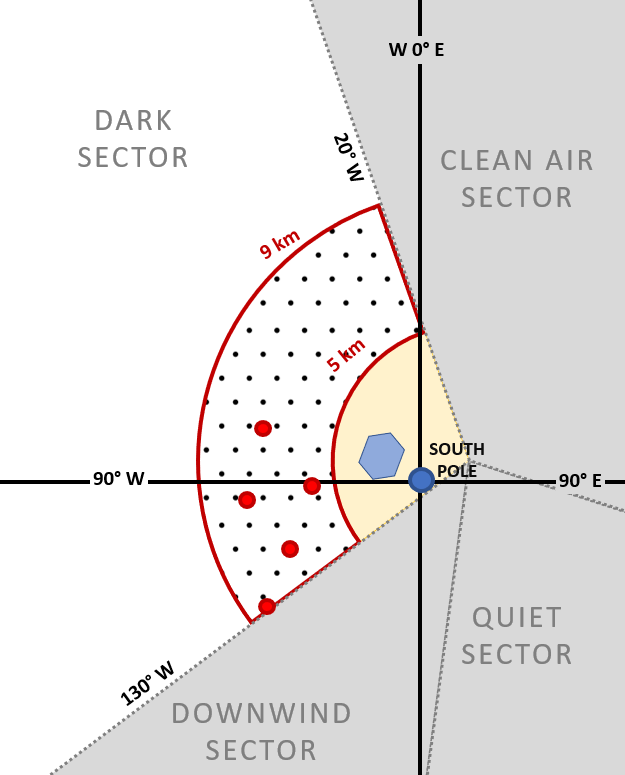}
    \caption{Layout of the proposed ARIA detector. The black dots represents the locations of the ARIA stations. The filled red circles show the positions of the ARA stations. The blue hexagon show the position of the IceCube detector.}
    \label{fig:array}
\end{figure}

ARIA consists of an array of 130 identical and independent neutrino stations, arranged in a hexagonal pattern and separated from each other by $\sim$\SI{1}{km}. The separation distance was determined by the requirement that $<$10\% of the neutrino events with E = \SI{e18}{eV} are observed by 2 or more stations. Though the observation of the same event by two independent stations provides a cross-check, it also reduces the sensitivity for a fixed number of stations. Since ARIA is a discovery instrument, it is useful to place ARIA stations as far apart as possible, keeping in mind deployment constraints. It requires less time to deploy stations over a smaller area. One possible geometric arrangement of the array is shown in Fig.~\ref{fig:array}.

\subsection{Science capabilities}
\begin{figure}[t]
    \centering
    \includegraphics[width=0.45\textwidth]{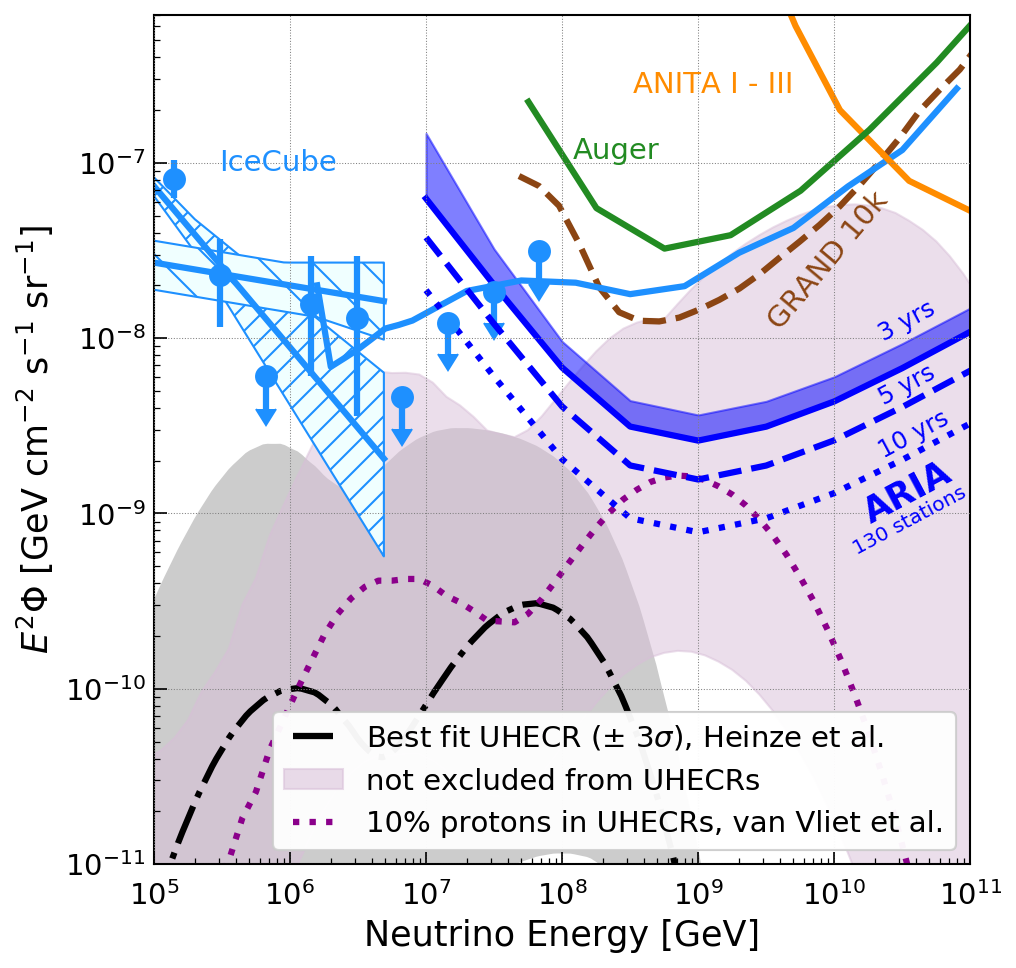}
    \includegraphics[width=0.45\textwidth]{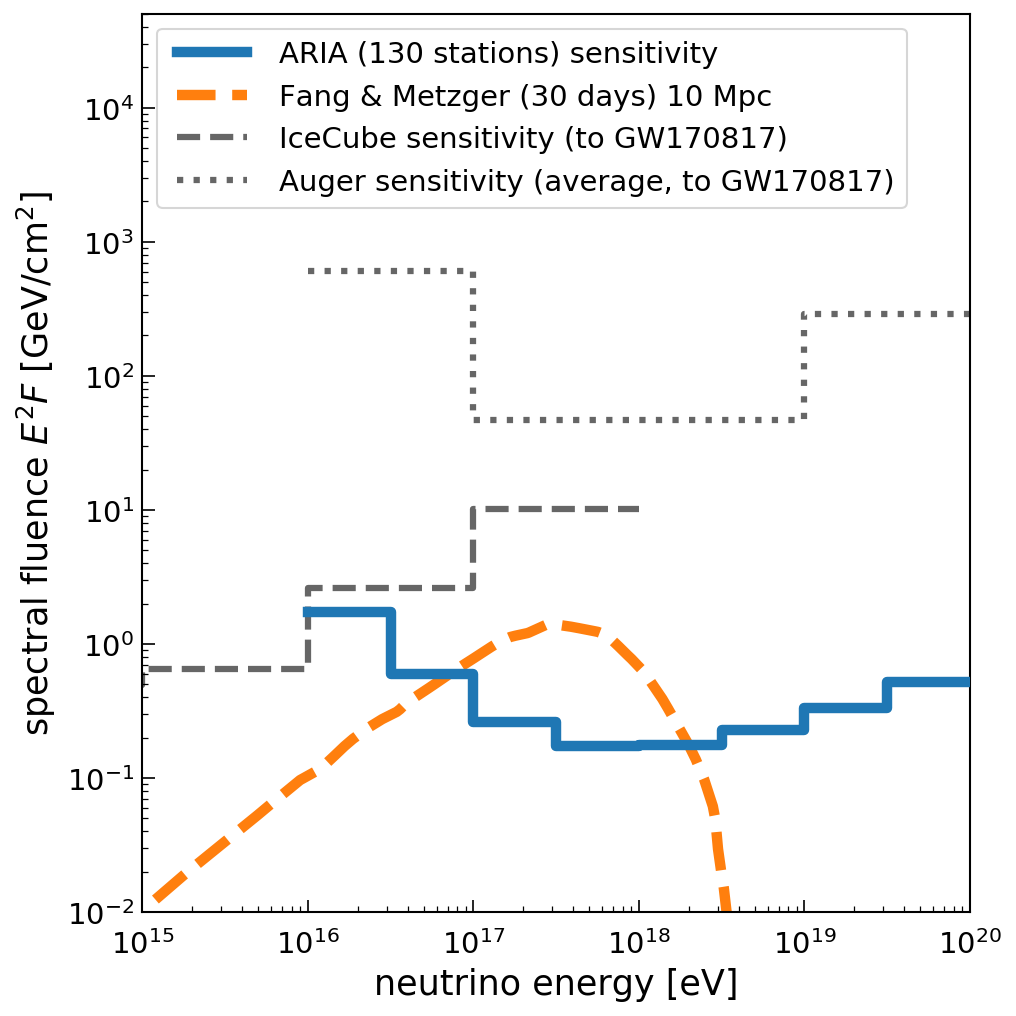}
    \caption{Expected sensitivity of the proposed ARIA detector in one-decade energy bins calculated using NuRadioMC. (left) ARIA sensitivity to an isotropic flux for 3, 5 and 10 years of operation assuming a uptime of 100\%. The shaded band of the 3 years expected limit represents uncertainties in the analysis efficiency. Also shown is the measured astrophysical neutrino flux from IceCube using the high-energy starting event (HESE) selection \citep{IceCubeFlux2017} and using a muon neutrino sample \citep{Haack2017}, limits from existing experiments (IceCube \citep{IceCubeFlux2018}, Auger \citep{AugerNeutrinoFlux2015} and Anita \citep{ANITA2018}), the expected sensitivity of the proposed GRAND10k detector \citep{GrandWhitePaper2018}, and several theoretical models. (right) ARIA sensitivity to transient sources if in field-of-view (solid blue line). Also shown is the sensitivity of current experiments \citep{NNFollowUp2017} and a theoretical model of a high-energy neutrino flux of a neutron-neutron star merger \citep{Fang2017a}.}
    \label{fig:sensitivity}
\end{figure}

The sensitivity of the ARIA detector is calculated using the state-of-the art NuRadioMC simulation code \citep{NuRadioMC} using the prediction of the Askaryan signal of \citep{AlvarezMuniz2000}. The expected sensitivity to both an isotropic neutrino flux as well as to transient sources is presented in Fig.~\ref{fig:sensitivity}. In 5 years of full-time operation, ARIA could limit the fraction of protons in the cosmic rays at the highest energies to 10\% or less for a standard choice of source evolution \citep{Vliet2019}, an important milestone\footnote{This is calculated as follows: We use the model of \citep{Vliet2019} of the cosmogenic neutrino flux for a source evolution parameter of $m = 3.4$, a spectral index of the injection spectrum of $\alpha = 2.5$, a cut-off rigidity of $R = \SI{100}{EeV}$, and a proton fraction of 10\% at $E = \SI{e19.6}{eV}$. We then calculate after how much lifetime, 2.44 neutrinos will be observed with the full ARIA detector. Assuming a non-observation and a zero background contribution, we can turn this into a 90\% Feldman-Cousins confidence upper limit on the proton fraction of cosmic rays.}. If the diffuse flux of high energy neutrinos discovered by IceCube continues to higher energies with a hard power law spectrum (dN/dE proportional to $E^{-2.1}$), ARIA will observe 12 events in 3 calendar years of full-time operation. It could be the first detection of a neutrino with energy \SI{> 3e7}{GeV}.

One potentially transformative method to understand the ultra-high-energy universe involves neutrino emission in coincidence with gravitational wave and/or electromagnetic emission. ARIA contributes to transient neutrino astronomy by virtue of it large instantaneous aperture, and broad field of view of $\Omega = \SI{2.1}{str}$ compared to many gamma-ray instruments. Fig.~\ref{fig:sensitivity} right shows the fluence limits of ARIA for a representative transient burst involving the merger of two neutron stars that was observed by gravitational wave detectors \citep{Abbott2017a}, and compares to current instruments. ARIA improves the sensitivity by more than a factor 10 compared to current instruments for E \SI{>e17}{GeV}.

\section{Conclusions}

The ARIANNA pilot array paves the way for a large-scale radio detector to discover cosmogenic neutrinos. The robust ARIANNA hardware has proven to work reliably in harsh Antarctic conditions. The data acquisition system with its precise time synchronization will enable the reconstruction of the telltale radio signatures of a high-energy neutrino interaction. 

The ARIANNA stations run completely autonomous with solar power through the summer and with a newly developed wind generator through most of the dark winter months. The development of a wind-power system that works at extreme cold temperatures is a great achievement as commercial systems fail quickly. In 2018, the first prototype survived the winter and was able to power an ARIANNA station for a substantial amount of time. This design is now being further improved to reach a higher wind yield. Hence, fully autonomous stations are a viable option for a large scale Askaryan detector at the South Pole. The autonomous nature of the design even allows for an installation at any deep ice location on the planet, an important goal to reach full sky coverage in the future. Furthermore, the real-time transmission of data via the Iridium satellite network and corresponding ultra-high energy neutrino alerts will contribute to the exiting multi-messenger effort to identify point sources. 

We demonstrated that radio signals originating from deep in the ice can be measured using a surface station via an in-situ calibration measurement, an important proof-of-concept. An ARIANNA station is able to reconstruct the predicted signal direction within a few degrees. This also shows that the ice properties and the signal propagation through the ice is well understood to correct for the bending of signal trajectories in the firn.

Neutrino signals can be identified with high efficiency and purity using a template matching technique that exploits the distinctive detector response to an impulsive signal. We used the more abundant cosmic-ray radio signals for an in-situ calibration and demonstration of the detector capabilities. We applied the template matching technique to obtain a pure sample of cosmic-ray events and found that the reconstructed polarization is in good agreement with the theoretical expectation.

Finally, guided by five years of successful operational experience with the ARIANNA pilot array, a large-scale high energy neutrino detector -- called ARIA -- was designed and proposed \citep{ARIA2018}. The station design was optimized for ice conditions at the South Pole. Neutrinos are distinguished from high rate background processes by several complementary channels, which is of utmost importance for a discovery instrument.
ARIA will have unprecedented sensitivity to high-energy neutrinos. It it sensitive to a proton fraction of as low as 10\% in the ultra-high energy cosmic-ray composition and will measure several astrophysical neutrinos per year if the flux observed by the IceCube detector continues to higher energies. 
Furthermore, the ARIA station is designed to have a good sensitivity to the neutrino direction and energy which enables multi-messenger astronomy at the highest neutrino energies.

\section*{Acknowledgements}
We are grateful to the U.S. National Science Foundation-Office of Polar Programs, the U.S.
National Science Foundation-Physics Division (grant NSF-1607719) and the U.S. Department of Energy. We thank generous support from the German Research Foundation (DFG), grant NE 2031/2-1 and GL 914/1-1, the Taiwan Ministry of Science and Technology. H. Bernhoff
acknowledge support from the Swedish Government strategic program Stand
Up for Energy. E. Unger acknowledge support from the Uppsala university
Vice-Chancellor's travel grant (sponsored by the Knut and Alice
Wallenberg Foundation) and the C.F. Liljewalch travel scholarships. D. Besson and A. Novikov acknowledge support from the MEPhI Academic Excellence Project (Contract No. 02.a03.21.0005) and the Megagrant 2013 program of Russia, via agreement 14.12.31.0006 from 24.06.2013

\section*{References}
\bibliography{mybibfile}
\end{document}